\documentclass[a4paper]{article}

\usepackage{latexsym}
\usepackage{amsmath}
\usepackage{amssymb}

\newtheorem{theorem}{Theorem}[section]
\newtheorem{prop}[theorem]{Proposition}
\newtheorem{lemma}[theorem]{Lemma}
\newtheorem{coro}[theorem]{Corollary}

\numberwithin{equation}{section}

\begin{document}

\def\Z{{\bf Z}} 
\def\x{{\bf x}} 
\def\del{\partial} 
\def\Lap{\bigtriangleup} 
\def\^{\wedge} 
\def\goes{\rightarrow}
\def\inv{^{-1}}

\def\reff#1{(\ref{#1})}
\def\vb#1{{\partial \over \partial #1}} % vector basis
\def\vbrow#1{{\partial/\partial #1}} % vector basis in row form
\def\Del#1#2{{\partial #1 \over \partial #2}}
\def\Dell#1#2{{\partial^2 #1 \over \partial {#2}^2}} \def\Dif#1#2{{d
    #1 \over d #2}} \def\Lie#1{ \mathcal{L}_{#1} } \def\diag#1{{\rm
    diag}(#1)} \def\abs#1{\left | #1 \right |} \def\rcp#1{{1\over #1}}
\def\paren#1{\left( #1 \right)} \def\brace#1{\left\{ #1 \right\}}
\def\bra#1{\left[ #1 \right]} \def\angl#1{\left\langle #1
  \right\rangle} 
\def\vector#1#2#3{\paren{\begin{array}{c} #1 \\ #2 \\
      #3 \end{array}}} 
\def\svector#1#2{\paren{\begin{array}{c} #1 \\
      #2 \end{array}}}
\def\matrix#1#2#3#4#5#6#7#8#9{ \left( \begin{array}{ccc}
        #1 & #2 & #3 \\ #4 & #5 & #6 \\ #7 & #8 & #9
        \end{array}  \right) }
\def\smatrix#1#2#3#4{ \left( \begin{array}{cc} #1 & #2 \\ #3 & #4
        \end{array}  \right) }

%Groups
\def\GL#1{{\rm GL}(#1)} \def\SL#1{{\rm SL}(#1)} \def\PSL#1{{\rm
    PSL}(#1)} \def\O#1{{\rm O}(#1)} \def\SO#1{{\rm SO}(#1)}
\def\IO#1{{\rm IO}(#1)} \def\ISO#1{{\rm ISO}(#1)} \def\U#1{{\rm
    U}(#1)} \def\SU#1{{\rm SU}(#1)}

%Abbrevs.
\def\Teich{{Teichm\"{u}ller}} \def\Poin{{Poincar\'{e}}}

\def\Gam{\mbox{$\Gamma$}} \def\d{{d}} \def\VII#1{\mbox{VII${}_{#1}$}}
\def\VI#1{\mbox{VI${}_{#1}$}}
\def\Isom{{\mathrm{Isom}}}

%Metrics
\def\hh{{h}}
\def\ggg{{\rm g}} 
% NOTE:
% Use ROMAN ``g'' to denote metric to distinguish from fundamental
% group elements
\def\uh#1#2{\hh^{#1#2}} 
\def\dh#1#2{\hh_{#1#2}}
\def\mh#1#2{\hh^{#1}{}_{#2}} 
\def\ug#1#2{\ggg^{#1#2}}
\def\dg#1#2{\ggg_{#1#2}} 
\def\uug#1#2{\tilde{\ggg}^{#1#2}}
\def\udg#1#2{\tilde{\ggg}_{#1#2}} 
\def\udh#1#2{\tilde{\hh}_{#1#2}}

\def\c#1{\chi_{#1}} \def\cc#1#2{\chi_{#1}{}^{#2}} \def\uc#1{\chi^{#1}}
\def\s#1{\sigma^{#1}} \def\ss#1#2{\sigma^{#1}{}_{#2}}
\def\om#1#2#3{\omega_{#1#2}{}^{#3}} \def\omd#1#2#3{\omega_{#1#2#3}}
\def\CC{C} \def\C#1#2#3{\CC^{#1}{}_{#2#3}} \def\Sig#1{\Sigma^{#1}}
\def\Sigg#1#2{\Sigma^{#1}{}_{#2}} \def\Chi#1{X_{#1}}
\def\Chii#1#2{X_{#1}{}^{#2}} \def\dug#1#2{g_{#1}{}^{#2}}
\def\dgm#1#2{\gamma_{#1#2}} \def\ugm#1#2{\gamma^{#1#2}}
\def\covsset{\mathcal{S}_2} \def\consset{\mathcal{S}^2} \def\X{X}
\def\dd#1#2{\frac{\del^2{#1}}{\del {#2}^2}}

\def\wa{\!\!\!\!&=&\!\!\!\!}  
\def\wb{\!\!\!\!&\equiv &\!\!\!\!}
\def\ws{&=} 
\def\nd{\noindent} 
\def\D{{\mathcal{D}}}

%Marks
\def\proofmark{\textsc{Proof.} } 
\def\defmark{{\bf Definition}\hspace{1em}} 
\def\defsmark{{\bf Definitions}\hspace{1em}} 
\def\notemark{{\bf Note.} }
\def\remarkmark{\textsc{Remark.} } 
\def\conventionmark{{\bf Convention.}}
\def\endofproofmark{\hfill\rule{.5em}{.5em}}

\def\R{\mathbf{R}} 

\def\N{\mathcal{N}}
\def\dq#1#2{q_{#1#2}}
\def\uq#1#2{q^{#1#2}}
\def\udq#1#2{\tilde{q}_{{#1}{#2}}}
\def\bdq#1#2{{\bar q}_{#1#2}}
\def\dqstd#1#2{\tilde{q}_{#1#2}^{(0)}}
\def\bdh#1#2{{\bar h}_{#1#2}}
\def\dhstd#1#2{\tilde{h}_{#1#2}^{(0)}}
\def\u{u}
\def\v{v}
\def\uv{\u\v}
\def\barA{\bar A}
\def\teiA{\u}
\def\teiB{\delta}
\def\teiC{\v}
\def\teiD{\uv}
\def\barphi{{\bar\phi}}
\def\signm{\varsigma}
\def\mass{\mathrm{m}_\Phi}
\def\am{\mathsf{L}} % ``angular momentum''
\def\AA{\mathcal{A}}

\def\bgam#1{\mathbf{\gamma}_{#1}}
\def\scss{\mathrm{(ss)}}
\def\notess{\mathrm{(0;ss)}}
\def\scren{\mathrm{(ren)}}

\def\harV{V}
\def\vEE#1#2#3{(\harV^{#1}_{#2})_{#3}}
\def\vE#1#2{\harV^{#1}_{#2}}
\def\vEEd#1#2#3{(\harV^{\prime #1}_{#2})_{#3}}
\def\vEd#1#2{\harV^{\prime #1}_{#2}}
\def\vEEdd#1#2#3{(\harV^{\prime\prime #1}_{#2})_{#3}}
\def\vEdd#1#2{\harV^{\prime\prime #1}_{#2}}

\def\b{\beta}
\def\G#1{G_{\mathrm{#1}}}

%% For comment
% \def\manbo#1{}
% \newbox{\manbobox}
% \newenvironment{manboo}%
%   {\begin{lrbox}{\manbobox}}%
%   {\end{lrbox}}

\def\AT{@aei.mpg.de}
\def\eml{Email}

\title{Harmonic Analysis of Linear Fields on the Nilgeometric
  Cosmological Model}
\author{{ Masayuki Tanimoto}
  \thanks{\eml: masayuki.tanimoto{\AT}} \\
  \textit{Max-Planck-Institut f\"ur Gravitationsphysik} \\
  \textit{(Albert-Einstein-Institut)} \\
  \textit{Am M\"uhlenberg 1, Golm 14476, Germany}}

\maketitle

\begin{abstract}
  To analyze linear field equations on a locally homogeneous spacetime
  by means of separation of variables, it is necessary to set up
  appropriate harmonics according to its symmetry group. In this
  paper, the harmonics are presented for a spatially compactified
  Bianchi II cosmological model --- the \textit{nilgeometric model}.
  Based on the group structure of the Bianchi II group (also known as
  the Heisenberg group) and the compactified spatial topology, the
  irreducible differential regular representations and the
  multiplicity of each irreducible representation, as well as the
  explicit form of the harmonics are all completely determined. They
  are also extended to vector harmonics. It is demonstrated that the
  Klein-Gordon and Maxwell equations actually reduce to systems of
  ODEs, with an asymptotic solution for a special case.
\end{abstract}

\section{Introduction}

A basic strategy to analyze linear field equations on a given
spacetime, like linear perturbation equations of Einstein's equation,
is to separate the equations using appropriate harmonics.  The
harmonics for a given manifold are in general determined by the
underlying symmetry group (isometry group) and the topology of the
manifold. The simplest example is that of a commutative group acting
on a flat manifold. If the manifold is compactified to, e.g., a torus,
then functions on the manifold are expanded in the form of usual
Fourier series. When the group is noncommutative, however, the
harmonics become much more complicated. The most familiar example in
homogeneous cosmology is the $\SU2$ (Bianchi IX) case
\cite{Bon71,Hu,JanJMP}, where one needs to use the spherical harmonics
(and their generalization to vector and tensor harmonics if necessary)
to separate field variables.  A notable nontrivial example is the
$H^2\times\R$ (Bianchi III) case \cite{TMY,Ta03} with compactified
three dimensional manifold. Since such a manifold is a direct product
of two submanifolds, a closed hyperbolic plane and a circle, the
harmonics are simply given by making products of those for the two
lower dimensional manifolds. Note that the Bianchi III belongs to
Class B \cite{EM}.  Separations of variables regarding locally
rotationally symmetric (LRS) Class A Bianchi types (and gravitational
perturbations) were discussed in \cite{Bon,JanEssay} without
compactification.

In this paper we consider the \textit{generic} (i.e., non-LRS) Bianchi
II model \textit{with} compactification. The Bianchi II type is one of
the class A types. The underlying symmetry group is the 3-dimensional
Heisenberg group $H_1$, which we refer to as the Bianchi II group
$G_\mathrm{II}$. The Bianchi II homogeneous manifolds correspond to
Thurston's nilgeometry \cite{Thu}.  We will refer to spatially
compactified Bianchi II spacetimes (i.e., ones obtained from Bianchi
II spatially homogeneous spacetimes by compactifying the homogeneous
spatial sections) as \textit{nilgeometric cosmological models} (to
distinguish from the conventional open models).

By considering a compactified (spatial) manifold we have the following
merits; (1) a compact Cauchy surface makes it natural to view the
field equations as an initial value problem in cosmological context,
(2) a finite spatial volume is physically reasonable, and (3)
functional analysis on a compact manifold is much easier and tractable
than one on an open manifold, since we can avoid many complexities
involved in continuous spectra and those for the convergence of
integrals (remember the contrast between Fourier series analysis on
the ``compactified space'' $S^1=\R^1/\Z$ and Fourier analysis on the
open space $\R$). 

We also mention that a motivation of considering compactified
manifolds also comes from recent evidences that the spatial
topology and global dynamical properties of solutions of Einstein's
equation are related in some ways. In particular, recent results
\cite{And,FM} suggest a very general picture of how an appropriately
conformally transformed spatial manifold evolves in time by the vacuum
Einstein equation, depending upon its topology. This picture motivates
us to study linear perturbation equations for locally homogeneous
solutions which have various spatial topologies.

Fortunately, although it is not as trivial as the torus
compactification of a flat manifold, the compactification of Bianchi
II manifolds (or Bianchi II type spacetimes) is not very difficult. We
will describe our compactification following \cite{KTH,TKH1,TKH2}.

In this paper we explore detailed properties for the scalar and vector
harmonics. We also demonstrate separation of variables for the
Klein-Gordon scalar field equation and the source-free Maxwell
equation. Although generalization to tensor harmonics is
straightforward we leave their explicit presentation to a subsequent
paper as well as a study of linear perturbations. This paper is
intended to lay a solid basis for exploring those more complicated
problems, or to be useful for many applications on the nilgeometric
spacetime model like quantum field analysis.

Some of the mathematical background assumed in this paper and some
related results obtained are the following.  As well known
(e.g.,\cite{Su}), the harmonics on a manifold $M$ on which a
transformation group $G$ acts are naturally obtained through
irreducible decomposition of a representation $T$, called the
\textit{regular representation}, of $G$ on $L^2(M)$. Let $g\in G$ and
let $f(x)\in L^2(M)$. The (right) regular representation $(T,L^2(M))$
is given by the homomorphism $T: g\goes T_g$, where $T_g$ is the right
translation map $T_g: f(x)\goes f(xg)$.\cite{note1} In fact, letting
$g,g'\in G$, we can see $T_gT_{g'}f(x)=(T_{g'}f)(xg)=
f(xgg')=T_{gg'}f(x)$, showing $T$ is a representation (homomorphism),
$T_gT_{g'}=T_{gg'}$. This representation is however \textit{not}
irreducible in general. The appropriate harmonics on $M$ are naturally
obtained through an irreducible decomposition of $(T,L^2(M))$.

In our context, the transformation group $G$ is the Bianchi II group
$G_\mathrm{II}$, which acts on $\tilde M=\R^3$ from the left simply
transitively, i.e., for arbitrary $p,q\in\tilde M$ there exist a
unique element $g\in G$ such that $gp=q$. Thanks to this property,
choosing an arbitrary fixed point $o\in \tilde M$, e.g., the
coordinate origin, one can identify the group $G_\mathrm{II}$ and the
manifold $\tilde M$, $G_\mathrm{II}=\tilde M$, by associating $go$
with $g$. With this identification, the manifold $\tilde M$ is also
the group $G_\mathrm{II}$, and therefore the right action of
$G_\mathrm{II}$ on $\tilde M$ is also naturally defined. We make
$\tilde M$ compact by identifying points by left action of a discrete
subgroup $A\subset G_\mathrm{II}$, $M=A\backslash \tilde M$. The
(right) regular representation $(T,L^2(A\backslash \tilde M))$ on this
space is indeed well defined, since keeping in mind the fact that we
can identify an arbitrary function on $A\backslash \tilde M$ with an
``automorphic function'' $f(x)$ on $\tilde M$ such that $f(x)=f(Ax)$,
we can confirm the consistency $T_gf(Ax)=f(Axg)=f(xg)=T_gf(x)$. This
shows the consistency we choose the ``right'' regular representation,
i.e., since we want to define the (Killing) symmetry of the manifold
with respect to the left action, and make a quotient by the left one,
the regular representation on the quotient should be the right one for
commutativity. The universal covering manifold $(\tilde
M,\tilde{q}^{(0)})$ with a standard left invariant metric
$\tilde{q}^{(0)}$ naturally defines a left invariant measure
$d\mu_0\equiv d\mu_\mathrm{L}$ on $\tilde M$, for which we can define
the natural inner product on $L^2(\tilde M)$. Since for the Bianchi II
group this measure is also right invariant $d\mu_\mathrm{L}\propto
d\mu_\mathrm{R}$, i.e., $G_\mathrm{II}$ is unimodular, the right
regular representation $(T,L^2(A\backslash\tilde M))$ with the inner
product is unitary; $\int_M |f(xg)|^2d\mu_0=\int_M |f(x)|^2d\mu_0$.

Our mode functions on $M$ will be denoted as
$\phi_{l,m,n_0}$ (or $\varphi_{l,m,n_0}$ on the spacetime $M\times
\R$) for generic modes with $m\neq0$. The index $m$ labels
inequivalent irreducible representations, while the index $n_0$ labels
ones in equivalent representations. For fixed $m$ and $n_0$, the
functions $\{\phi_l\}_{l=0}^\infty$ work as a set of basis functions
for the irreducible representation space specified by $m$ and
$n_0$. Each single $\phi_l$ spans the eigenspace of an operator
denoted as $\am^2$, which is like a total angular momentum operator.

For the purpose of separation of variables, the most important
relations are those for the differential representation, which is a
linear transformation acting on the representation space spanned by mode
functions. Those relations are written in terms of group-invariant
differential operators denoted as $\c I$ ($I=1\sim 3$).  (See
Eqs.\reff{eq:relc123} and \reff{eq:c-phinote}.)  Indeed, since
group-invariant field equations like the Klein-Gordon equation can be
written with no explicit coordinate dependences if it is written with
the invariant operators, these relations are found to provide the key
to separate the equations.

The formulas for the differential representation however do not
provide complete information about the representation. For example,
they do not tell whether the representation specified by $m$ does
exist in $(T,L^2(M))$, or how many copies of an equivalent
representation exist in $(T,L^2(M))$. To see how the representation
$(T,L^2(M))$ is decomposed to irreducible representations we need to
find all the appropriate mode functions on $M$. In this paper the
universal covering manifold $\tilde M$ (with a group invariant
standard metric) is compactified to a circle bundle over the torus.
The regular representation $(T,L^2(M))$ is completely reducible, and
as a result of finding of the mode functions on the given topology of
$M$ we find the following:
\begin{theorem}
  \label{th:1}
  Let $M$ be the circle bundle over the torus with Euler class $e=1$
  (see Eq.\reff{eq:pi1} for its fundamental group),
  and let $(T,L^2(M))$ be the regular representation of the Bianchi II
  group $G_\mathrm{II}$. Then, it holds that
  \begin{equation}
    T=\paren{\bigoplus_{m\in\Z\backslash\{0\}}|m|T_m}\bigoplus
    \paren{ \bigoplus_{k_1,k_2\in\Z}
    \mathbf{1}_{k_1,k_2} },
  \end{equation}
  where $T_m$ is an infinite dimensional irreducible representation,
  $\mathbf{1}_{k_1,k_2}$ is a one dimensional irreducible
  representation, and the coefficient $|m|$ stands for the
  multiplicity in $T_m$. $\Z\backslash\{0\}$ represents nonzero
  integers.
\end{theorem}
(See \cite{Fol}, \S10 and \S11 and references therein for related
mathematical works.) This decomposition expresses the completeness of
the harmonics we construct. In particular, this decomposition does not
depend on the \Teich\ (or moduli) parameters of $M$. From the
Stone--von Neumann theorem \cite{Fol,Tay}, $T_m$ is equivalent to a
corresponding Schr\"odinger representation.

Construction of vector (or tensor) harmonics is not difficult on one
hand. However the important point is to divide each irreducible space
of vectors into subspaces such that each subspace is invariant under
the action of the operator $\am^2$. This feature is necessary to
obtain decoupled systems of ODEs when the background spacetime has an
additional symmetry. We define three kinds of vector harmonics, two of
which have this property.

The plan of the paper is as follows. In the next section we describe
the background solution and also account for some basic facts. Section
\ref{ssec:II-2} sets up some basic eigenvalues used to label mode
functions, based on the compactification of the spatial manifold. In
\S.\ref{sec:properties} we make algebraic discussions to derive the
$\chi$-relations. Section \ref{sec:mf} is devoted to construction of
the mode functions on the spatial manifold. In section
\ref{sec:gensptm} the mode functions constructed on the spatial
manifold are generalized to those on the spacetime. While
\S\S.\ref{sec:properties}--\ref{sec:gensptm} deal with the generic
modes, \S.\ref{sec:U1-sym} deals with the exceptional modes, the
$U(1)$-symmetric modes, which complete all possible (scalar) modes.
The results so far are applied to the Klein-Gordon equation and the
reduced ODEs are explicitly given, with an asymptotic solution for a
special case, in \S.\ref{sec:KG}. In \S.\ref{sec:vh} we develop the
vector harmonics. Section \ref{sec:max} is devoted to an application
to Maxwell's equation. The final section is devoted to conclusion.

This paper is a full account, with much generalizations and
development, of the subject outlined in \S 3 of \cite{Ta03}. Although
most notations remain the same, one of the changes is that a
quotient by the left action is now written $A\backslash\tilde M$
instead of $\tilde M/A$ to make clear which action is used.
We employ the abstract index notation \cite{Wa} and use leading Latin
letters $a,b,\cdots$ to denote abstract indices for vectors and
tensors in \S\S.\ref{sec:KG}--\ref{sec:max}. In the other sections
however we write them without abstract indices. We often drop the
tensor product symbol and write, e.g., $(\s1)^2$ instead of
$\s1\otimes\s1$. Beware that since vectors are also used as
differential operators, products of them like $(\c1)^2$ can stand for
second order derivatives like $\c1\c1$ or tensor products like
$\c1\otimes\c1$, depending upon the quantity considered.

%%%%%%%%%%%%%%%%
\section{The background solution}
\label{ssec:II-1}

Our background solution is specified by the following: (1) it is a
solution of the vacuum Einstein equation, (2) it is spatially locally
homogeneous of Bianchi II type, and (3) its spatial manifold is
compact without boundary, in other words, \textit{closed}. (An
explicit topology will be chosen later.)

A Bianchi type II solution is characterized by the fact that the
solution (or the universal cover of it) is invariant under the action
of the \textit{Bianchi II group} $\G{II}$, which is a
three-dimensional nilpotent (e.g., \cite{Tay}) Lie group. The group
multiplication is given by
\begin{equation}
  (a,b,c)(a',b',c')=(a+a',b+b',c+c'+ab'),
\end{equation}
for $(a,b,c),(a',b',c')\in G_\mathrm{II}$. (\textbf{Note}: To save
space we try to express components of group elements in a row form as
above, but a column form is also equally used when it is more
convenient.)

Let $\tilde M=\R^3$ be the simply-connected open manifold with
coordinates $(x,y,z)$. We can define the group action on this manifold
identifying the group manifold $\G{II}$ with $\tilde M$.  The
left-action is therefore expressed as
\begin{equation}
  \label{eq:axmulti}
  (a,b,c)(x,y,z)=(a+x,b+y,c+z+ay),
\end{equation}
where $\mathbf a=(a,b,c)\in\G{II}$, and $\mathbf x=(x,y,z)\in \tilde
M(\simeq G_\mathrm{II})$. Let $\xi_I$ $(I=1,2,3)$ be the generators of
the one-parameter subgroups $(a,0,0)$, $(0,b,0)$ and $(0,0,c)\in
G_\mathrm{II}$. It is easy to find they are expressed
\begin{equation}
  \xi_1=\vb x+y\vb z,\quad \xi_2=\vb y, \quad\xi_3=\vb z.
\end{equation}
Similarly, the generators of the right actions and their dual
one-forms are given by
\begin{equation}
  \label{eq:invbasis}
\begin{split}
      \chi_1 = \vb x,\quad \chi_2=\vb y+x\vb z, \quad\chi_3=\vb z, \\
      \s1 = dx,\quad \s2=dy, \quad \s3=dz-x dy.
\end{split}
\end{equation}
These vectors $\c I$ and one-forms $\s I$
($I=1\sim3$) are called the \textit{invariant vectors or one-forms} of
$\G{II}$, since they are left invariant;
\begin{equation}
  \Lie{\xi_I}\c J=[\xi_I,\c J]=0=\Lie{\xi_I}\s J, \quad I,J=1\sim3,
\end{equation}
where $\Lie{\xi_I}$ is the Lie derivative with respect to $\xi_I$.
The invariant vectors satisfy the following commutation relations:
\begin{equation}
  \label{eq:comGIIchi}
  [\c1,\c2]=\c3,\quad [\c2,\c3]=0,\quad [\c3,\c1]=0.
\end{equation}
The vectors $\xi_I$ ($I=1,2,3$) are \textit{Killing vectors} for the
metric of the form $\tilde{q}=\udq IJ\s I\otimes \s J$ with the components
$\udq IJ$ being constants. Riemannian manifold $(\tilde M,\tilde{q})$ is
called \textit{homogeneous}, since $G_\mathrm{II}$ acts transitively
on it as its isometry group.

A homogeneous metric is called \textit{locally rotationally symmetric
  (LRS)} if it has a fourth independent Killing vector $\xi_4$.
Bianchi type II LRS metrics are given by the metrics of the form
$\tilde q^{(\mathrm{LRS})}=\udq11
((\s1)^2+(\s2)^2)+\udq33(\s3)^2$, since such a metric has an
additional Killing vector, given by
\begin{equation}
  \xi_4=-y\vb x+x\vb y+\rcp2(x^2-y^2)\vb z.
\end{equation}
This Killing vector generates the following one-parameter isometries
$s_\theta=e^{\theta\xi_4}$ for the metric $\tilde
q^{(\mathrm{LRS})}$:
\begin{equation}
  s_\theta: \; \vector xyz\goes
  \svector{R_\theta\svector{x}{y}}{z+\zeta_\theta(x,y)},
\end{equation}
where $(x,y,z)\in \tilde M$, $R_\theta$ is the rotation matrix
$R_\theta=\smatrix{\cos\theta}{-\sin\theta}{\sin\theta}{\cos\theta}$,
and
\begin{equation}
  \zeta_\theta(x,y)\equiv
  \rcp2((x^2-y^2)\cos\theta-2xy\sin\theta)\sin\theta.
\end{equation}
An LRS manifold $X=(\tilde M,\tilde q^{(\mathrm{LRS})})$ has as a
result a four dimensional isometry group $\Isom X$. Let $\Isom_0X$ be
its identity component, i.e., the component connected to the identity.
An element $\mathbf{\alpha}\in\Isom_0X$ can be uniquely expressed as
the composite $(\alpha_1,\alpha_2,\alpha_3)\circ s_\theta$ for a
choice of $(\alpha_1,\alpha_2,\alpha_3)\in \G{II}$ and $s_\theta$.

The one-parameter diffeomorphism $s_\theta$ plays an important role even
when the metric is not LRS. It forms a one-parameter subgroup of the
automorphism group of $\G{II}$, which induces on the cotangent space a
rotation of the invariant one-forms:
\begin{equation}
  \label{eq:s*}
  s_{\theta *}: \; \vector{\s1}{\s2}{\s3}\goes
  \svector{R_\theta\svector{\s1}{\s2}}{\s3}.
\end{equation}
The induced map $s_{\theta}^*$ on tangent space acts on the invariant
vectors $\c I$ the same way with replacement $R_\theta \goes
R_{-\theta}$ above. The significance of the automorphisms of Bianchi
groups was first fully recognized by Jantzen \cite{Jan}. Maps
$s_{\theta *}$ or $s_{\theta}^*$ will be used in this paper in several
contexts.

We can obtain the spacetime metric for the conventional Bianchi
cosmology assuming that all the components with respect to the
invariant frame $(dt,\s I)$ formed by the invariant one-forms and the
timelike basis $dt$ are functions of time $t$ only. The vacuum
Bianchi II solution $\tilde{\ggg}$ was first obtained by Taub \cite{Taub}.
We write that metric in the following form using our invariant
one-forms \reff{eq:invbasis}:
\begin{equation}
    \tilde{\ggg}=-N^2(t)dt^2+q_1(t)(\s1)^2+q_2(t)(\s2)^2+q_3(t)(\s3)^2,
\end{equation}
where
\begin{equation}
  \label{eq:AQ}
  N^2= 1+\b^2t^{4p_3},\quad
  q_1=t^{2p_1}N^2,\quad
  q_2=t^{2p_2}N^2,\quad
  q_3=16p_3^2\b^2t^{2p_3}/N^2.
\end{equation}
Parameters $p_i(i=1,2,3)$ and $\b$ are constants such that $\b>0$,
$p_3\neq0$, and
\begin{equation}
  \Sigma p_i=\Sigma p_i^2=1.
\end{equation}

When $p_1=p_2$, the solution is LRS. Although there exist two possible
such cases $(p_1,p_2,p_3)= (0,0,1)$ (case I LRS) and $(2/3,2/3,-1/3)$
(case II LRS), these two solutions represent equivalent one-parameter
solutions. In fact, we can check that the case I LRS solution with
$\b=\b_{\mathrm{I}}$ is isometric to the case II LRS solution with
$\b=\b_{\mathrm{II}}=3^{-2/3}\b_{\mathrm{I}}^{-1/3}$.  When we are
interested in an LRS solution, the case II LRS solution may be
preferable, since the time coordinate $t$ in this solution approaches
the proper time $\tau$ at future infinity. This will make comparisons
with other models like the Bianchi type III \cite{TMY,Ta03} more
straightforward.

It is also worth pointing out that a solution $\tilde{\ggg}$ with
$(p_1,p_2,p_3)$ is isometric to another solution $\tilde{\ggg}'$ with
$p_1$ and $p_2$ swapped. In fact, it is at once using Eq.\reff{eq:s*}
to see that $\tilde{\ggg}'$ is the metric induced by $s_{\pi/2}$;
$\tilde{\ggg}'=s_{\pi/2 *}\tilde{\ggg}$. We can therefore without loss
of generality assume, e.g., $p_1\leq p_2$.

We denote the conventional solution described so far as $(\tilde
M\times\R,\tilde{\ggg})$, and call it the \textit{universal covering
  solution}. On the other hand our spatially closed solution, denoted
$(M\times\R,\ggg)$, is obtained introducing a spatial
compactification with it. We express the solution as
\begin{equation}
  \label{eq:bIIsol}
  (M\times\R,\ggg)=\Gamma\backslash (\tilde M\times\R,\tilde{\ggg}),
\end{equation}
using an appropriate discrete subgroup $\Gamma$ of $\G{II}$ which acts
spatially from the left on the solution. The metric $\ggg$ here is
the one induced from $\tilde{\ggg}$. ($\ggg$ and $\tilde{\ggg}$ are
therefore locally isometric to each other.) While there are infinitely
many possible compactifications (i.e., spatial topologies), we for
definiteness specify the spatial manifold $M$ to be ``the circle
bundle over the 2-torus with Euler class $e=1$.''  (See, e.g.,
\cite{Hem,Sco}. In general, a closed Bianchi II manifold is a
\textit{Seifert fiber space} over a Euclidean orbifold.)  The
fundamental group can be represented in the standard notation as
\begin{equation}
  \label{eq:pi1}
  \pi_1(M)=\angl{g_1,g_2,g_3;[g_1,g_2]=g_3,[g_1,g_3]=1,[g_2,g_3]=1},
\end{equation}
where the brackets stand for group commutators, $[a,b]\equiv aba\inv
b\inv$. The procedure for the actual compactification is described in
the next section.  The resulting spatially compactified generalization
was first constructed and discussed in \cite{TKH1}. Note that as a
result of the compactification the spatial manifold specified by
$t=\mathrm{constant}$ is now \textit{locally homogeneous}
(e.g.,\cite{Sco}), and the spacetime solution is said to be
\textit{spatially locally homogeneous}.

As shown in \cite{TKH1}, $\Gamma$ contains four free parameters (see
Eq.\reff{eq:Gamrep}). Our spatially closed solution \reff{eq:bIIsol}
therefore forms a \textit{six} parameter solution (since the
universal cover has as we have seen \textit{two} independent
parameters, $\b$ and, e.g., $p_3$).

%%%%%%%%%%%%%%%%%%%
\section{Compactification and eigenvalues}
\label{ssec:II-2}

To proceed, we need to describe the compactification of the spatial
manifold and thereby define some eigenvalues.

Let us first describe the canonical way of expressing a Bianchi type
II locally homogeneous manifold $(M,q)$. Let $\tilde{q}^{(0)}$ be the
standard metric given by
\begin{equation}
  \tilde{q}^{(0)}=(\s1)^2+(\s2)^2+(\s3)^2
\end{equation}
and let 
\begin{equation}
 \N\equiv(\tilde M,e^{2\alpha}\tilde{q}^{(0)}) 
\end{equation}
be the \textit{standard conformal manifold}, with $e^{2\alpha}$ being
a constant conformal factor. Then \cite{KTH}, the manifold $(M,\dq
ab)$ can be expressed as a quotient of such a standard conformal
manifold (with an appropriate choice of the factor $e^{2\alpha}$),
\begin{equation}
    (M,q)=A\backslash \N,
\end{equation}
where $A$ is an appropriate discrete subgroup of the isometry group of
$\N$, $A\subset \Isom\N$.

The subgroup $A$ must be isomorphic to the fundamental group
$\pi_1(M)$ given by \reff{eq:pi1}. (In fact, $M$ is a Haken
manifold \cite{Hem}.) This means that $A$ must be an embedding of
$\pi_1(M)$ into the isometry group $\Isom\N$. Let
$\mathbf{a}_i\in\Isom\N$ ($i=1,2,3$) be the image of $\pi_1$-generator
$g_i$ by such an embedding. Following the procedure shown in
\cite{KTH}\cite{note2},
we find that it is possible to parameterize them in the following way:
\begin{equation}
  \mathbf{a}_1=(\u,\delta,0),\quad
  \mathbf{a}_2=(0,2\pi\v,0),\quad
  \mathbf{a}_3=(0,0,2\pi\uv).
\end{equation}
Here, $\mathbf{a}_i\in \G{II}\subset\Isom\N$.\cite{note3}  We denote
$A=\brace{\mathbf{a}_1,\mathbf{a}_2,\mathbf{a}_3}$. The three real
parameters $\u$, $\v$, and $\delta$ are called the \textit{\Teich\ 
  parameters of geometric structure} for the locally homogeneous
3-manifold $(M,q)$.

We construct the harmonics on $(M,q)$ by two steps; first we do
the construction on a covering manifold, denoted $(\bar M,\bar{q})$,
and then superpose appropriate subset of the harmonics to obtain those
on $(M,q)$. The \textit{auxiliary manifold} $(\bar M,\bar{q})$ is
simply defined by removing $\mathbf{a}_1$ from $A$, i.e., $(\bar
M,\bar{q})=\barA\backslash\N$, where
$\barA=\brace{\mathbf{a}_2,\mathbf{a}_3}$.  $\barA$ is a commutative
subgroup of $A$, and therefore $\bar M$ is homeomorphic to the much
tractable manifold $T^2\times\R$. In fact, we can easily see that it
is each $x=\mathrm{constant}$ plane in $\N$ that is compactified to a
2-torus.

Now, let us introduce some important operators. We define the ``total
angular momentum-like'' operator
\begin{equation}
  \label{eq:defam^2}
  \am^2\equiv (\Lie{\c1})^2+ (\Lie{\c2})^2,
\end{equation}
using the Lie derivatives $\Lie{\c1}$ and $\Lie{\c2}$. Note that when
acting on a scalar, it becomes a simple form
\begin{equation}
  \am^2= (\c1)^2+ (\c2)^2,
\end{equation}
which also coincides with the Laplacian $\Lap_0$ with respect to the
standard metric $\tilde{q}^{(0)}$, up to square of $\c3$:
\begin{equation}
  \label{eq:Lap0}
 \Lap_0=\am^2+(\c3)^2,
\end{equation}
when acting on a scalar. Here, $\c I$ ($I=1,2,3$) are regarded as
differential operators. It is quite important to recognize that
operators $\c I$ are well defined not only on the universal cover
$\tilde M$ but also on the compactified manifold $M=A\backslash \tilde
M$. In other words, the induced vector fields $\pi^* \c I$ on $M$ for
the covering map $\pi: \; \tilde M\goes M=A\backslash \tilde M$ is
well defined because of the invariance of $\c I$ under the action of
$A\subset\G{II}$.  (For simplicity we do not explicitly write $\pi^*$,
and identify $\pi^*\c I$ and $\c I$.)  Since $\bar A$ is also a
subgroup of $\G{II}$, $\c I$ are well defined on $\bar M$, also.  The
globally-defined invariant operators $\c I$ work as the fundamental
derivative operators, since any group-invariant field equations become
independent of coordinates when they are written with $\c I$. This
coordinate-free property of the field equations is necessary to be
able to reduce the field equation to ordinary differential equations.

The operator $\c3$, called the \textit{fiber generator}, has a special
importance, since it commutes with all the invariant operators $\c I$.
In other words, $\c3$ is the \textit{center} of the Bianchi II
algebra. As a direct consequence of Schur's lemma, such an operator
must be diagonalized to obtain an irreducible representation of the
regular representation mentioned in Introduction.

The operator $\am^2=(\c1)^2+(\c2)^2$ commutes with $\c 3$, since $\c3$
is the center. We may therefore be able to diagonalize our mode
functions with respect to both $\c3$ and $\am^2$. Also, consider
another operator $\xi_2=\del/\del y$, which we can find commutes with
both $\am^2$ and $\c3$, so we may diagonalize the mode functions with
respect to $\xi_2$, also. This operator however is \textit{not} well
defined on $M$, but on $\bar M$. This is the reason we consider the
auxiliary manifold $\bar M$. The existence of $\xi_2$ is important to
make it possible to perform separation of variables for the eigenvalue
equation for $\am^2$ (see below).

Let $i\mu$ and $i\nu$ be eigenvalues for the operators $\c3$ and
$\xi_2$:
\begin{equation}
  \label{eq:c3xi2}
  \c3 \bar\phi=i\mu\bar\phi,\quad \xi_2 \bar\phi=i\nu\bar\phi.
\end{equation}
The function $\bar\phi$ is supposed to be an appropriate mode
function on $\bar M$. Also, we define $\lambda$ by
$\am^2\bar\phi=-\lambda^2\bar\phi$. The solution of these equations
is given by
\begin{equation}
  \bar\phi=X(x)e^{i\mu z}e^{i\nu y},
\end{equation}
where the function $X(x)$ is a solution of the following ``harmonic
oscillator Schr\"odinger equation'':
\begin{equation}
  \label{eq:X}
  \frac{d^2\X}{dx^2}+({\lambda^2-({\mu x+\nu})^2})\X=0.
\end{equation}
For $\bar\phi$ to be well defined on $\bar M$, it must be an
automorphic function such that $\bar\phi(\bar
A\x)=\bar\phi(\x)$.\cite{note4}  From
this condition we find
\begin{equation}
  \label{eq:defmunu}
  \begin{split}
  \mu &= \mu(m)= m/(\uv), \quad m\in\Z, \\
  \nu &= \nu(n)= n/\v, \quad n\in \Z.
  \end{split}
\end{equation}
We call $\mu$, $m$, $\nu$, and $n$, respectively, the \textit{fiber
  eigenvalue}, \textit{fiber index}, \textit{auxiliary eigenvalue},
and \textit{auxiliary index}. We call $\lambda^2$ the \textit{total
  eigenvalue}. The spectrum of $\lambda^2$ is determined in the next
section.

%%%%%%%%%%%%%%%%%%
\section{Irreducible differential representations}
\label{sec:properties}

It is not difficult to determine the irreducible representations of
the regular representation in their differential form, i.e., the
differential representations of the Bianchi II algebra.  In fact, it
will be found that this procedure is similar to the one in determining
quantum states of the harmonic oscillator, since $\am^2$ (or the
scalar Laplacian $\Lap_0$) has essentially the same algebraic
structure as that of the Hamiltonian of the oscillator.

As mentioned in the previous section since the fiber generator $\c3$
must be a constant when acting on an irreducible subspace, the fiber
index $m\in\Z$ does not change values in this space. We therefore
assume that $m$ is fixed throughout this section. It may be helpful to
bear in mind that as for the correspondence to quantum mechanics, the
fiber eigenvalue $\mu=m/\teiD$ corresponds to the Planck constant $h$,
while $\c1$ and $\c2$ correspond, respectively, to the position $x$
and momentum $p$ operators, as in $[x,p]=ih\Leftrightarrow
[\c1,\c2]=\c3=i\mu$.  Remember however that our representation space
is $L^2(M)$ instead of $L^2(\R)$.

In this section we deal with the generic $m\neq0$ case. The
exceptional $m=0$ case will be discussed in \S.\ref{sec:U1-sym}.

Let $\phi$ be an eigenfunction on $M$ for the operators $\c 3$ and
$\am^2$, i.e., $\c3\phi=i\mu\phi$, and $\am^2\phi=-\lambda^2\phi$. It
is helpful to introduce a symbol signifying the sign of the fiber
eigenvalue, which allows us to discuss both $m\gtrless 0$ cases
simultaneously; we define
\begin{equation}
  \signm\equiv \mathrm{sign}(m)=\mathrm{sign}(\mu). \quad (m\neq0)
\end{equation}
Let us then define
\begin{equation}
  \label{eq:As}
  \AA_1\equiv \frac1{\sqrt{2}}(\chi_1+\signm i\chi_2),\quad
  \AA_2\equiv \frac1{\sqrt{2}}(\chi_1-\signm i\chi_2),\quad
  \AA_3\equiv -\signm i\chi_3.
\end{equation}
Then, we immediately find the following commutation relations
\begin{equation}
  \label{eq:comAL}
  [\am^2,\AA_1]=\signm 2i\AA_1\chi_3,\quad
  [\am^2,\AA_2]=-\signm 2i\AA_2\chi_3.
\end{equation}
This means that $\AA_1$ and $\AA_2$ are, respectively, a \textit{raising
  and lowering operator} for the total eigenvalue $\lambda^2$. In
fact, since
\begin{equation}
  \am^2 \AA_1\phi=([\am^2,\AA_1]+\AA_1\am^2)\phi
  =  (\signm 2i\AA_1\chi_3+\AA_1\am^2)\phi
  = -(2|\mu|+\lambda^2)\AA_1\phi,
\end{equation}
$\AA_1\phi$ is an eigenfunction for $\lambda'{}^2=\lambda^2+2|\mu|$.
Similarly, $\AA_2\phi$ is an eigenfunction for
$\lambda'{}^2=\lambda^2-2|\mu|$.

Taking into account the fact that $\AA_1$ and $\AA_2$ change the
eigenvalue $\lambda^2$ by $\pm 2|\mu(m)|$, we can without loss of
generality assume the form of spectrum as
\begin{equation}
  \label{eq:l}
  \lambda^2=|\mu|(2l+c_m),
\end{equation}
where
\begin{equation}
  \label{eq:l:range}
  l=0,1,\cdots.
\end{equation}
The value for $l=0$, $\lambda^2=\lambda^2_0\equiv|\mu|c_m$,
corresponds to the smallest one for given $m$, which must exist
because minus the Laplacian $-\Lap_0=-\am^2-(\c3)^2=-(\am^2+\mu^2)$
can have only nonnegative eigenvalues. We call $l$ the \textit{spin
  index}.

At this point we know the eigenmode is specified by the pair of
integers $(l,m)$, so the corresponding eigenfunction can be expressed
with these labels $\phi_{l,m}$. As we remarked since the value of $m$
does not change in an irreducible space, we drop $m$ and write
$\phi_{l}$ for simplicity.

Let us write down the whole relations we have as
\begin{equation}
  \label{eq:rel:LapA312}
  \begin{split}
    \am^2\phi_{l}&= -|\mu|(2l+c_m)\phi_{l}, \\
    \AA_3\phi_{l}&= |\mu| \phi_{l}, \\
    \AA_1\phi_{l}&= \alpha_{l} \phi_{l+1}, \\
    \AA_2\phi_{l}&= \beta_{l} \phi_{l-1},
  \end{split}
\end{equation}
where we have introduced unknown constants $\alpha_{l}$ and
$\beta_{l}$, which possibly depend on $l$.

Note the following identity that can be easily checked by a direct
computation:
\begin{equation}
  \label{eq:Lapidentity}
  \am^2= 2\AA_1\AA_2- \AA_3.
\end{equation}
Using Eqs.\reff{eq:rel:LapA312}, this implies $ -|\mu|(2l+c_m)=
2\alpha_{l-1}\beta_{l}- |\mu|$, i.e.,
\begin{equation}
  \label{eq:relab}
  \alpha_{l-1}\beta_{l}=-|\mu|(l+\frac{c_m-1}{2}).
\end{equation}
Because of the arbitrariness of constant multipliers for the
eigenfunctions $\phi_l$, we may set $\alpha_{l}$ or $\beta_{l}$
arbitrarily, but once it is set, the other is constrained from this
relation. We take\cite{note5}
\begin{equation}
  \label{eq:atmp}
    \alpha_{l}= -\sqrt{|\mu|}, \quad
  \beta_{l}= \sqrt{|\mu|}(l+\frac{c_m-1}{2}).
\end{equation}
These do satisfy Eq.\reff{eq:relab}.  Then, since we defined $l$ so
that $l=0$ gives, for given $m\neq0$, the smallest eigenvalue of
$-\am^2$, we should have
\begin{equation}
\label{eq:annbyA2}
\AA_2\phi_{0}=\beta_{0}\phi_{-1}=
\sqrt{|\mu|}(\frac{c_m-1}{2})\phi_{-1}=0,
\end{equation}
implying
\begin{equation}
  \label{eq:cm}
  c_m=1.
\end{equation}
So, now we have
\begin{equation}
  \label{eq:a}
  \alpha_{l}= -\sqrt{|\mu|},\quad 
  \beta_{l} = \sqrt{|\mu|}\,l.
\end{equation}
Gathering Eqs.\reff{eq:As},\reff{eq:rel:LapA312},
and \reff{eq:a}, we arrive at the following set of
relations:
\begin{equation}
  \label{eq:relc123}
  \begin{split}
  \chi_1\phi_{l}&= -\sqrt{\frac
  {|\mu|} 2}\paren{\phi_{l+1}-l\phi_{l-1}}, \\
  \chi_2\phi_{l}&= \signm i\sqrt{\frac
  {|\mu|} 2}\paren{\phi_{l+1}+l\phi_{l-1}}, \\
  \chi_3\phi_{l}&= i\mu\phi_{l}, \\
  l &= 0,1,\cdots .
  \end{split}
\end{equation}
In particular,
\begin{equation}
  \label{eq:deflambda^2_l}
  \am^2\phi_l=-\lambda^2_l\phi_l, \quad \lambda^2_l\equiv |\mu|(2l+1).
\end{equation}

Now, we have found the following. A ``ground state'' $\phi_0$ is
determined as a solution for the two equations
\begin{equation}
  \label{eq:A320pos}
  \AA_3\phi_0= |\mu| \phi_0,\quad \AA_2\phi_0=0.
\end{equation}
Note that the function $\phi_0$ obtained this way is automatically
an eigenfunction of $\am^2$ as seen from the identity
\reff{eq:Lapidentity}.  The excited states $\phi_l$ are determined by
successively multiplying the raising operator $(-1/\sqrt{|\mu|})\AA_1$,
i.e.,
\begin{equation}
  \label{eq:annipos}
  \phi_l=\paren{-\frac{\AA_1}{\sqrt{|\mu|}}}^l\phi_0.
\end{equation}
The space spanned by these functions, $L^2_m(M)\equiv
\{\sum_{l=0}^\infty a_l\phi_l| a_l\in\mathbf{C}\}\cap L^2(M)$, gives
an irreducible subspace of $L^2(M)$. In other words, the restriction
of the regular representation $T$ to $L^2_m(M)$, denoted as
$(T_m,L^2_m(M))$, gives an irreducible representation. The
differential representation $(dT_m,L^2_m(M))$ is given by
Eqs.\reff{eq:relc123}, which will be repeatedly used to separate the
field equations. For convenience, we call these relations the
$\chi$-\textit{relations}.

%%%%%%%%%%%%%%%%%%%
\section{Mode functions on the compactified manifold}
\label{sec:mf}

In this section we solve the eigenvalue equations for the mode
functions $\phi_l$ under the appropriate automorphic conditions. As a
result we find how many equivalent copies of the irreducible
representation $T_m$ are contained in $T$, in other words, the
multiplicity of $T_m$ is determined. The explicit form of $\phi_l$
itself is also of great interest. In this section we continue to
assume $m\neq0$.

One of the possible procedures to find explicit form of $\phi_l$ is to
solve the equations \reff{eq:A320pos} to find $\phi_0$ and compute
successive differential operations in Eq.\reff{eq:annipos} to find
general $\phi_l$.  Another procedure is to directly solve the
eigenvalue equation $\am^2\phi_l=-\lambda^2\phi_l$ for general spin
index $l$. While both ways are possible, we take the latter, which
provides quicker way of identifying the solutions with known
functions.

As remarked in \S.\ref{ssec:II-2}, let us find the mode functions on
$\bar M$ first. Note that Eq.\reff{eq:X} becomes (attaching index $l$
to $X$)
\begin{equation}
  \frac{d^2X_l}{d\zeta^2}+\paren{l+\rcp2-\frac{\zeta^2}4}X_l=0,
\end{equation}
if we define
\begin{equation}
  \label{eq:zeta}
  \zeta=\signm\sqrt{\frac{2}{\abs{\mu}}}(\mu x+\nu).
\end{equation}
Independent solutions to the above equation are given by
$D_l(\zeta)$ and $D_{-l-1}(i\zeta)$, where $D_l(\zeta)$ is the Weber
parabolic cylinder function. When $l$ is zero or a positive integer,
$D_l(\zeta)$ can be expressed using the Hermite polynomial
$H_l(\zeta)$;
\begin{equation}
  D_l(\zeta)=e^{-\rcp4\zeta^2}H_l(\zeta).
\end{equation}
Our convention for the Hermite polynomial is $H_l(\zeta)=
(-1)^le^{\rcp2\zeta^2}({d^l}/{d\zeta^l})e^{-\rcp2\zeta^2}$.
  
Since $\phi_0$ must be annihilated by $\AA_2$ the appropriate choice is
found to be $D_l(\zeta)$, i.e., we must take
$X_l=\mathrm{constant}\times D_l(\zeta)$.  In fact, the equation
$\AA_2\bar\phi_{0}=0$ together with the separation form
$\bar\phi_{l}=X_{l}(x)e^{i\nu y}e^{i\mu z}$, implies
$(d/d\zeta+(1/2)\zeta)X_{0}=0$, with the solution being
$X_{0}=\mathrm{constant}\times e^{-(1/4)\zeta^2}$.  This coincides
with the one claimed for $l=0$. (Conversely, as we will see, functions
$\phi_{l}$ constructed using these $X_l$ can satisfy the desired
relations \reff{eq:relc123} for all $l$, which justifies our choice.)

The mode functions on $\bar M$ are therefore, attaching indices $m$
and $n$, given by
\begin{equation}
  \barphi_{l,m,n}(\x)=C_{l}
  D_l(\pm\sqrt{\frac{2}{\abs{\mu}}}(\mu x+\nu))e^{i\mu z}e^{i\nu y},
\end{equation}
where $C_{l}$ are constants and $\mu$ and $\nu$ are defined in
Eqs.\reff{eq:defmunu}.

The constants $C_l$ are determined by requiring that the functions
$\barphi_{l,m,n}(\x)$ obey the $\chi$-relations \reff{eq:relc123}.
Using the widely known formulas
\begin{equation}
  \begin{split}
    D_l'(\zeta) &= -\rcp2(D_{l+1}(\zeta)-lD_{l-1}(\zeta)), \\
    \zeta D_l(\zeta) &= D_{l+1}(\zeta)+lD_{l-1}(\zeta),
  \end{split}
\end{equation}
we can easily find $C_{l+1}=C_{l}$, i.e., they are constants that
do not depend on $l$;
\begin{equation}
  C_l=C.
\end{equation}
(Actually, this is the reason we chose Eqs.\reff{eq:a}.) The constant
$C$ may be determined by a normalization of the square integral on
$M$. (See below.)

The mode functions on $M$ are, as mentioned, expressed as an infinite
sum of these eigenfunctions on $\bar M$.  Remember that they must be
invariant under the action of
$A=\{\mathbf{a}_1,\mathbf{a}_2,\mathbf{a}_3\}$, and the functions
$\bar\phi_{l,m,n}$ are already invariant under $\bar
A=\{\mathbf{a}_2,\mathbf{a}_3\}$. We therefore want to make a linear
combination of $\bar\phi_{l,m,n}$ so that it is invariant under
$\mathbf{a}_1$. Recalling the multiplication rule \reff{eq:axmulti},
we find the following transformation law (cf. \cite{Ta03}, Eq.(3.15))
\begin{equation}
  \bar\phi_{l,m,n}(\mathbf{a}_1\x)= 
  e^{i\frac{\teiB}{\teiC}n}\bar\phi_{l,m,n+m}(\x).
\end{equation}
From this we can see that the following function $\phi_{l,m,n_0}$,
defined as an infinite sum, is actually invariant under the action of
$\mathbf{a}_1$ (cf. \cite{Ta03}, Theorem 3.1), i.e.,
$\phi_{l,m,n_0}(\mathbf{a}_1\x)=\phi_{l,m,n_0}(\x)$ for
\begin{equation}
  \label{eq:philm2}
  \phi_{l,m,n_0}(\mathbf{x}) =
  \sum_{k=-\infty}^{\infty} 
  e^{ i{\teiB}(n_0k+m\frac{k(k-1)}{2})}
  \barphi_{l,m,n_0+m k}(\mathbf{x}),
\end{equation}
where $ l = 0,1,\cdots, \infty, \quad |m| = 1,2,\cdots, \infty, \quad
n_0 = 0,1,\cdots, |m|-1 $. The sum is convergent at any point $\x$.
It is easy to see that since the functions $\barphi_{l,m,n}(\x)$
satisfy the relations \reff{eq:relc123}, so do $\phi_{l,m,n_0}(\x)$.
These functions are therefore the right mode functions on
$M=A\backslash\tilde M$.

As a result of the compactification, the index $n_0$ for the mode
functions on $M$ is now bounded by $|m|$.  $|m|$ is the
\textit{multiplicity} of the modes specified by the same $l$ and $m$.
Since for each $n_0$ the functions $\phi_{l,m}$ span an irreducible
subspace of $L^2(M)$, $|m|$ is also the multiplicity of the
irreducible representation $T_m$ contained in $T$.

We can summarize the results as follows.
\begin{theorem}
  There exist $|m|$ different sets of mode functions
  $\{\phi_{l}(\x)\}_{l=0}^\infty$ that satisfy the relations
  \reff{eq:relc123} on the compactified manifold $(M,\dq
  ab)=A\backslash\N$ (with $M$ being the $S^1$-bundle over the 2-torus
  with Euler class $e=1$).
\end{theorem}

\bigskip

Let us discuss how the mode functions can be normalized. We define the
inner product in $L^2(M)$ as
\begin{equation}
  \label{eq:innerM}
  (f,g)\equiv \int_Mfg^* d\mu_0,
\end{equation}
where $g^*$ is the complex conjugate of $g$,
$d\mu_0=\s1\^\s2\^\s3=dxdydz$ is the standard invariant measure.
We want to determine the square norm
\begin{equation}
  N_l\equiv (\phi_l,\phi_l).
\end{equation}
We first observe the following.
\begin{lemma}
  The invariant operators $\c I$ ($I=1\sim 3$) in $L^2(M)$ are
  anti-selfadjoint, $\c I^\dagger =-\c I$.
\end{lemma}

\proofmark Since
\begin{equation}
  (\c I f,g)= \int_M (\c I(fg^*)-f\c Ig^*)d\mu_0
  = \mathcal{I}_I-(f,\c I g),
\end{equation}
we need to show $\mathcal{I}_I\equiv \int_M \c I(fg^*)d\mu_0=0$. In
fact, when, e.g., $I=1$, we can show $\mathcal{I}_1=\int_M
\c1(fg^*)\s1\^\s2\^\s3=\int_M d(fg^*\s2\^\s3)$, which is from Stoke's
theorem $\int_{\del M}fg^*\s2\^\s3=0$. Here, we have used the
identity
\begin{equation}
 df=(\c1f)\s1+(\c2f)\s2+(\c3f)\s3,
\end{equation}
which is valid for an arbitrary function $f$ on $M$,
and also used the relation $d(\s2\^\s3)=0$, which is confirmed from
the definition \reff{eq:invbasis}. The other cases $I=2,3$ are the
same, since $d(\s1\^\s3)=d(\s1\^\s2)=0$.  \endofproofmark

\remarkmark Operators $(1/i)\c I$ ($I=1\sim 3$) are selfadjoint.

\begin{coro}
  \label{coro:AdaggA}
  In $L^2(M)$, $\AA_1^\dagger=-\AA_2$.
\end{coro}

\proofmark $\AA_1^\dagger=2^{-1/2}(\c1+\signm i\c2)^\dagger=
2^{-1/2}(\c1^\dagger-\signm i\c2^\dagger)=
-2^{-1/2}(\c1-\signm i\c2)=-\AA_2$. \endofproofmark

Returning to the issue of $N_l$, consider
$N_{l+1}=(\phi_{l+1},\phi_{l+1})$. When $m\neq 0$, from
Eqs.\reff{eq:rel:LapA312} and Corollary \ref{coro:AdaggA} we have
\begin{equation}
  \label{eq:N_l+1proptoN_l}
  \begin{split}
    N_{l+1} &= \frac{1}{\alpha_l\alpha_l^*}(\AA_1\phi_l,\AA_1\phi_l) \\
    &= \frac{1}{\alpha_l\alpha_l^*}(\phi_l,\AA_1^\dagger \AA_1\phi_l) \\
    &= \frac{-1}{\alpha_l\alpha_l^*}(\phi_l,\AA_2 \AA_1\phi_l) \\
    &= \frac{-1}{\alpha_l\alpha_l^*}(\phi_l,\beta_{l+1}\alpha_l\phi_l) \\
    &= \frac{-\beta_{l+1}^*}{\alpha_l}N_l.
  \end{split}
\end{equation}
Substituting our choice \reff{eq:a} of $\alpha_l$ and $\beta_l$ we have
\begin{equation}
  N_{l+1}=(l+1)N_l.
\end{equation}
Taking $N_0=1$ we conclude
\begin{equation}
  N_l=l!.
\end{equation}
Now, we have the following.
\begin{theorem}
  Suppose that $\phi_l$ are mode functions on $(M,\dq
  ab)=A\backslash\N$ such that they satisfy the relations
  \reff{eq:relc123}. Multiplying the same constant normalization
  factor $C$ to all $\phi_l$, $\phi_l\goes C\phi_l$, does not change
  those relations. By choosing $C$ appropriately, we can make the
  normalization
  \begin{equation}
    \label{eq:thenorm}
    (\phi_{l,m,n_0},\phi_{l',m',n_0'})=
    l!\, \delta_{ll'}\delta_{mm'}\delta_{n_0n_0'}.
  \end{equation}
  hold.
\end{theorem}

\proofmark The orthogonality for $m$ and $m'$ is apparent from the
fact that $\mu(m)$ is the eigenvalue of the selfadjoint operator
$(1/i)\c3$. The orthogonality for $n_0$ and $n_0'$ comes from the
orthogonality among the mode functions on $\bar M$:
\begin{equation}
  (\barphi_{l,m,n},\barphi_{l,m,n'})_{\bar M}\equiv
  \int_{\bar M}\barphi_{l,m,n}\barphi_{l,m,n'}^*d\mu_0= 0,
  \quad (n\neq n'),
\end{equation}
which is also apparent from the fact that the operator $(1/i)\xi_2$,
of which eigenvalues are $\nu(n)$, is selfadjoint on $L^2(\bar M)$.
Observing that when $n_0\neq n_0'$, $\phi_{l,m,n_0}$ and
$\phi_{l,m,n_0'}$ are linear combinations in different sets
$\{\barphi_{l,m,n}\}_{n\in \mathcal{N}_1}$ and
$\{\barphi_{l,m,n}\}_{n\in \mathcal{N}_2}$,
$\mathcal{N}_1\cap\mathcal{N}_2=\emptyset$, we can easily see
\begin{equation}
  \int_{\bar M}\phi_{l,m,n_0}\phi_{l,m,n_0'}^*d\mu_0= 0, \quad
  (n_0\neq n_0')
\end{equation}
which in tern implies the orthogonality for $n_0$ and $n_0'$ in
$L^2(M)$. The other part has already been proven. \endofproofmark

\remarkmark The constant $C$ does not depend on $l$, but can depend on
$m$ and $n_0$, so we may write $C=C_{m,n_0}$.

\bigskip

Apparently, if we define
\begin{equation}
  \phi_{l,m,n_0}^\mathrm{(n)}\equiv\frac{1}{\sqrt{l!}}\,\phi_{l,m,n_0},
\end{equation}
they become orthonormal to each other:
\begin{equation}
  (\phi_{l,m,n_0}^\mathrm{(n)},\phi_{l',m',n_0'}^\mathrm{(n)})
  = \delta_{ll'}\delta_{mm'}\delta_{n_0n_0'}.
\end{equation}
As seen from Eqs.\reff{eq:N_l+1proptoN_l} and \reff{eq:relab} this
corresponds to choosing
\begin{equation}
  \alpha_l^\mathrm{(n)}= e^{i\Theta_l}\sqrt{|\mu|(l+1)},\quad
  \beta_l^\mathrm{(n)}= -e^{-i\Theta_l}\sqrt{|\mu|l},
\end{equation}
where $\Theta_l$ is an arbitrary phase factor, which we may want to
take zero, $\Theta_l=0$. Substituting $\alpha_l=\alpha_l^\mathrm{(n)}$
and $\beta_l=\beta_l^\mathrm{(n)}$ into Eqs.\reff{eq:rel:LapA312} we
have another version of $\chi$-relations for
$\phi_{l,m,n_0}^\mathrm{(n)}$, which have the most direct
correspondence to the usual relations between the quantum states of
the harmonic oscillator. As mentioned however we employ the
unnormalized $\phi_{l}$ in this paper for the convenience of
computations.

%%%%%%%%%%%%%%%%%%
\section{Further transformation}
\label{sec:gensptm}

The mode functions $\phi_{l,m,n_0}(\x)$ shown in the previous section
are \textit{not} well defined on the general spacetime solution
$(M\times\R,\ggg)=\Gamma\backslash (\tilde M\times\R,\tilde{\ggg})$,
because of the fact $A\neq\Gamma$. Remember that $\Gamma\subset
G_\mathrm{II}$ is a four-parameter embedding, while $A$ is a
three-parameter one. This incompatibility means that we cannot
identify the coordinates $\x$ in $\N$ with the spatial coordinates
$\x$ in $(\tilde M\times\R,\tilde{\ggg})$.\cite{note6} Although it is
expected that an appropriate diffeomorphism can make the mode
functions well defined on the spacetime, such a diffeomorphism can
affect the $\chi$-relations. In the following, we show by explicit
computations that this is the case but a further renormalization makes
the mode functions retain the original $\chi$-relations.

As shown in \cite{TKH1}, the covering group $\Gamma$ can be
parameterized as
\begin{equation}
  \label{eq:Gamrep}
 \Gamma=\{\bgam1,\bgam2,\bgam3\}=\left\{
   \svector{R_{-\theta}\svector{\teiA}{\teiB}}{0},
   \svector{R_{-\theta}\svector{0}{2\pi\teiC}}{0},
   \vector{0}{0}{2\pi\teiA\teiC} \right\},
\end{equation}
where $\bgam i\in\G{II}$, and $\theta$, $\teiA$, $\teiB$, and $\teiC$
are real parameters.\cite{note7}\cite{note8} Let
$(M,q)=\Gamma\backslash (\tilde M,\udq ab)$ be the \textit{spatial
  section} of the solution, where $\tilde{q}$ is the spatial part of
$\tilde{\ggg}$. For convenience of considering diffeomorphisms between
the spatial universal cover $(\tilde M,\udq ab)$ and the standard
conformal manifold $\N=(\tilde M,e^{2\alpha}\tilde{q}^{(0)})$, let us
distinguish the latter manifold (without metric structure) by denoting
$\tilde M'$.  This distinction is meant to be helpful just to keep
track of the direction of the diffeomorphisms we consider.

Let $\psi$ be a diffeomorphism:
\begin{equation}
  \psi: \tilde M\goes\tilde M'
\end{equation}
such that
\begin{equation}
  \Gamma=\psi\inv\circ A\circ\psi.
\end{equation}
Then, the induced function of $\phi_{l,m,n_0}(\x)$ will be invariant
under $\Gamma$, since so is $\phi_{l,m,n_0}(\x)$ under $A$.

We can find $\psi$ in $\Isom\N$, which is explicitly given by
\begin{equation}
  \label{eq:defpsi}
  \psi=\mathbf{b}\circ s_{\theta},
\end{equation}
where $\mathbf{b}\in\G{II}$ is
\begin{equation}
  \mathbf{b}=(\pi \teiC\cos\theta\sin\theta,
  \teiA\inv(-\zeta_{-\theta}(\teiA,\teiB)+
  \pi\teiB\teiC\sin\theta\cos\theta),0).
\end{equation}
We have, for simplicity, chosen the third component of $\mathbf{b}$ as
zero, though it can be an arbitrary constant. The induced vectors
$\psi^{*}\c I$ on $\tilde M'$ becomes a linear transformation of $\c
I$, due to the property that elements in $\Isom\N$ are automorphisms
of $\G{II}$. Note that when acting on the vectors $\c I$, the induced
map $\psi^*=\mathbf{b}^*\circ s_{\theta}^*$ becomes the same as
$s_{\theta}^*$, since $\c I$ are by definition invariant under the
induced map $\mathbf{b}^*$ for $\mathbf{b}\in \G{II}$. Therefore from
(the vector version of) Eq.\reff{eq:s*}, we have
\begin{equation}
  \label{eq:psipush}
  \psi^{*}:\; \vector{\c1}{\c2}{\c3}\goes
  \vector{\psi^{*}\c1}{\psi^{*}\c2}{\psi^{*}\c3}
  =\svector{R_{-\theta}\svector{\c1}{\c2}}{\c3}.
\end{equation}
Let $\phi_l^\scss(\x)\equiv(\psi_*\phi_l)(\x)=\phi_l\circ\psi(\x)$,
where $\x\in\tilde M$. (Superscript $(\mathrm{ss})$ stands for
``spatial section''.)  Then,
\begin{equation}
  \begin{split}
    \c1\phi_l^\scss &= \c1(\psi_*\phi_l)
    = (\psi^*\c1)\phi_l\circ\psi \\
    &= (\cos\theta\c1+\sin\theta\c2)\phi_l\circ\psi \\
%    &= (\cos\theta(\c1\phi_l)+\sin\theta(\c2\phi_l))\circ\psi \\
    &= \sqrt{\frac{|\mu|}2}
    \paren{-e^{-i\theta}\phi_{l+1}+e^{i\theta}l\phi_{l-1}}\circ\psi \\
    &= \sqrt{\frac{|\mu|}2}
    \paren{-e^{-i\theta}\phi_{l+1}^\scss+e^{i\theta}l\phi_{l-1}^\scss},
  \end{split}
\end{equation}
where we have used the relations \reff{eq:relc123}. Similarly, we
obtain
\begin{equation}
    \c2\phi_l^\scss=
    \signm i\sqrt{\frac{|\mu|}2}
    \paren{e^{-i\theta}\phi_{l+1}^\scss+e^{i\theta}l\phi_{l-1}^\scss},
\end{equation}
and $\c3\phi_l^\scss=i\mu\phi_l^\scss$. These relations are different
from the original $\chi$-relations \reff{eq:relc123} unless $\theta$
is a multiple of $2\pi$.

However, it is possible to renormalize $\phi_l^\scss$ so that the
original $\chi$-relations are recovered. It is straightforward
to check that
\begin{equation}
  \varphi_{l}
  =e^{i\theta l}\phi_{l}^\scss
\end{equation}
gives such a renormalized function. Thus, we have found that the
functions $\varphi_{l}=e^{i\theta l}\phi_{l}\circ\psi$ with $\phi_l$
given by Eq.\reff{eq:philm2} and $\psi$ being Eq.\reff{eq:defpsi}, are
served as the right mode functions on the spatial section (and
therefore on the spacetime) that satisfy the relations
\reff{eq:relc123}.  This provides a direct
proof of the following:
\begin{theorem}
  \label{th:gensptm}
  There exist $|m|$ different sets of time-independent mode functions
  $\{\varphi_{l,m}(\x)\}_{l=0}^\infty$ on the spatially closed Bianchi
  II solution $(M\times\R,\ggg)=\Gamma\backslash (\tilde
  M\times\R,\tilde{\ggg})$ such that they satisfy the relations
  \reff{eq:relc123}.
\end{theorem}

%%%%%%%%%%%%%%%%%%
\section{$U(1)$-symmetric modes}
\label{sec:U1-sym}

Let us, for completeness, consider the modes with the fiber index $m$
being zero. We call these modes $U(1)$-\textit{symmetric}, since they
are constant along the $U(1)$ ($\simeq S^1$) fibers.

Let $\phi$ be an eigenfunction for $m=0$, i.e.,
$\c3\phi=(\del\phi/\del z)=0$. This in turn implies that $\c1$ and
$\c2$ are commutative when acting on $\phi$,
\begin{equation}
  [\c1,\c2]\phi=0,
\end{equation}
since $[\c1,\c2]=\c3$. Due to this property, the harmonics describing
the $m=0$ subspace of $L^2(M)$ become the usual Fourier expansion on a
torus. (Thus, together with the results for the generic modes we
obtain Theorem \ref{th:1}.) In the following we explicitly determine
the spectrum of the eigenvalues for the operators $\c1$ and $\c2$ in
terms of the spacetime moduli parameters $\teiA$, $\teiB$, $\teiC$,
and $\theta$.

Let us first work on the 3-manifold $A\backslash\N$ (not on the
spacetime manifold) as we did for the generic ($m\neq 0$) case. Taking
the form of $A$ into account, we label the mode functions with the the
following equations
\begin{equation}
  \label{eq:eigeneqs0}
  \begin{split}
    (\teiA\c1+\teiB\c2)\phi &= 2\pi i k_1\phi, \\
    \teiC\c2\phi &= i k_2\phi,
  \end{split}
\end{equation}
where the eigenvalues $k_1$ and $k_2$ are to be used as labels. Let
us therefore write the solution of these equations as
$\phi=\phi_{k_1,k_2}^{(0)}(\x)=\phi_{k_1,k_2}^{(0)}(x,y)$, which is
given by
\begin{equation}
  \label{eq:phik1k2}
  \phi_{k_1,k_2}^{(0)}=\mathrm{constant}\times e^{\frac i\teiA(2\pi
  k_1x+\frac{k_2}{\teiC}(-\teiB x+\teiA y))}.
\end{equation}
Since, recalling the rule \reff{eq:axmulti}, 
\begin{equation}
  \begin{split}
  \phi_{k_1,k_2}^{(0)}(\mathbf{a}_1 \x)
  &= \phi_{k_1,k_2}^{(0)}(x+\teiA,y+\teiB)=
  \phi_{k_1,k_2}^{(0)}(\x)e^{2\pi ik_1},  \\
  \phi_{k_1,k_2}^{(0)}(\mathbf{a}_2 \x)
  &= \phi_{k_1,k_2}^{(0)}(x,y+2\pi\teiC)=
  \phi_{k_1,k_2}^{(0)}(\x)e^{2\pi ik_2},
  \end{split}
\end{equation}
we find
\begin{equation}
  k_1\in\Z,\quad k_2\in\Z,
\end{equation}
for $\phi_{k_1,k_2}^{(0)}$ to be well defined on $M=A\backslash\tilde
M$. The remaining condition
$\phi_{k_1,k_2}^{(0)}(\mathbf{a}_3 \x)=\phi_{k_1,k_2}^{(0)}(\x)$
is trivial.

Therefore from Eqs.\reff{eq:eigeneqs0} the $\chi$-relations for
the $U(1)$-symmetric modes are
\begin{equation}
  \label{eq:relcphi0123}
  \begin{split}
  \c1\phi^{(0)}_{k_1,k_2} &= i K_1(k_1,k_2)\,\phi^{(0)}_{k_1,k_2} \\
  \c2\phi^{(0)}_{k_1,k_2} &= i K_2(k_2)\,\phi^{(0)}_{k_1,k_2} \\
  \c3\phi^{(0)}_{k_1,k_2} &= 0,
  \end{split}
\end{equation}
where
\begin{equation}
  \begin{split}
    K_1(k_1,k_2) &\equiv
    \frac{1}{\teiA}(2\pi k_1-\frac{\teiB}{\teiC}k_2), \\
    K_2(k_2) &\equiv \frac{k_2}{\teiC}.
  \end{split}
\end{equation}

To extend the mode functions on the spacetime manifold
$\Gamma\backslash (\tilde M\times\R,\tilde{\ggg})$, we need to apply the
diffeomorphism $\psi$ defined in Eq.\reff{eq:defpsi} again, and as a
result the spectrum of the eigenvalues are altered. (Contrary to the
generic case, any renormalizations of the resulting mode functions do
not affect the $\chi$-relations.)

As in the previous section, let us define
$\varphi^{(0)}_{k_1,k_2}\equiv\phi^{(0)}_{k_1,k_2}\circ\psi$. Then,
from Eq.\reff{eq:psipush}, we have, e.g.,
\begin{equation}
  \begin{split}
    \c1\varphi^{(0)}_{k_1,k_2} &= 
    (\psi^*\c1)\phi^{(0)}_{k_1,k_2}\circ\psi \\
    &= (\cos\theta\c1+\sin\theta\c2)\phi^{(0)}_{k_1,k_2}\circ\psi \\
    &= i(\cos\theta K_1+\sin\theta K_2)\varphi^{(0)}_{k_1,k_2}.
  \end{split}
\end{equation}
A similar result is also obtained for $\c2\varphi^{(0)}_{k_1,k_2}$. We
write the final form of the relations as follows.
\begin{equation}
  \label{eq:c-phinote}
\begin{split}
  \c1\varphi_{k_1,k_2}^{(0)} &= 
  i \kappa_1(k_1,k_2)\varphi_{k_1,k_2}^{(0)}, \\
  \c2\varphi_{k_1,k_2}^{(0)} &= 
  i \kappa_2(k_1,k_2)\varphi_{k_1,k_2}^{(0)}, \\
  \c3\varphi_{k_1,k_2}^{(0)} &= 0,\quad k_1\in\Z, \; k_2\in\Z,
\end{split}
\end{equation}
where
\begin{equation}
  \begin{split}
    \kappa_1(k_1,k_2) &\equiv \cos\theta K_1(k_1,k_2)+\sin\theta
    K_2(k_2), \\
    \kappa_2(k_1,k_2) &\equiv -\sin\theta K_1(k_1,k_2)+\cos\theta
    K_2(k_2).
  \end{split}
\end{equation}
Now, we have:
\begin{theorem}
  There exist time-independent mode functions
  $\varphi_{k_1,k_2}^{(0)}(\x)$, $k_1,k_2\in\Z$, on the spatially
  closed Bianchi II solution $(M\times\R,\ggg)=\Gamma\backslash
  (\tilde M\times\R,\tilde{\ggg})$ such that they satisfy the relations
  \reff{eq:c-phinote}.
\end{theorem}

%%%%%%%%%%%%%%%%%%
\section{Application to the Klein-Gordon equation}
\label{sec:KG}

As an example, let us consider the Klein-Gordon equation
\begin{equation}
  (\ug ab\nabla_a\nabla_b-\mass^2)\Phi=0,
\end{equation}
where $\mass\geq 0$ is the mass of the field $\Phi$. $\nabla_a$ is the
covariant derivative operator associated with the spacetime metric
$\dg ab$. It is straightforward to see that this equation on our
background can be expressed, using the invariant operators $\c I$, as
\begin{equation}
  \label{eq:waveII}
  \paren{ \frac{-1}{\sqrt{-\ggg}}\vb t\paren{\sqrt{-\ggg}N^{-2}\vb t}
  + \Lap_q -\mass^2}
\Phi=0,
\end{equation}
where $\Lap_q$ is the Laplacian with respect to the spatial metric
$\dq ab$;
\begin{equation}
  \label{eq:L_q}
  \Lap_q=q_1\inv(\chi_1)^2+q_2\inv(\chi_2)^2+q_3\inv(\chi_3)^2,
\end{equation}
and $\sqrt{-\ggg}\equiv \sqrt{-\det{\dg ab}}=4|p_3|\b t N^2$.

Let us consider a generic irreducible component of $\Phi$, i.e.,
$\Phi=\Phi_{m,n_0}$, $m\neq0$.  We can expand this component as
\begin{equation}
  \Phi(t,\mathbf{x})=\sum_{l=0}^\infty
  a_{l}(t)\varphi_{l}(\mathbf{x}),
\end{equation}
where $\varphi_{l}=\varphi_{l,m,n_0}$ are the mode functions mentioned
in Theorem \ref{th:gensptm}.

From the relations \reff{eq:relc123}, we have
\begin{equation}
  \label{eq:cdouble}
  \begin{split}
    (\c1)^2\varphi_{l} &=
    \frac{|\mu|}{2}
    \paren{ \varphi_{l+2}-(2l+1)\varphi_l+l(l-1)\varphi_{l-2} },
    \\
    (\c2)^2\varphi_{l} &=
    -\frac{|\mu|}{2}
    \paren{ \varphi_{l+2}+(2l+1)\varphi_l+l(l-1)\varphi_{l-2}
    }, \\
    (\c3)^2\varphi_{l} &= -\mu^2\varphi_l,
  \end{split}
\end{equation}
from which we immediately obtain the following wave
equations for $a_l(t)$:
\begin{equation}
  \label{eq:wIIfora}
  \ddot a_l+\rcp t\dot a_l+ Z(t) a_l= I(t;a_{l-2},a_{l+2}),
\end{equation}
where
\begin{equation}
  \begin{split}
  Z(t) & \equiv
    \frac{\mu^2}{16(p_3)^2\b^2}(1+\b^2t^{4p_3})^2t^{-2p_3}
    +\mass^2(1+\beta^2t^{4p_3}) \\
    & \quad +\frac{2l+1}2 |\mu| (t^{-2p_1}+t^{-2p_2}),
  \end{split}
\end{equation}
with the inhomogeneous term $I$ being
\begin{equation}
  I(t;a_{l-2},a_{l+2})\equiv
  \frac{|\mu|}{2} (t^{-2p_1}-t^{-2p_2})(a_{l-2}+(l+2)(l+1)a_{l+2}).
\end{equation}
(In $I(t;a_{l-2},a_{l+2})$, $a_{l-2}$ should be regarded zero when
$l=0$ and $1$.)

Note that the inhomogeneous term $I$ introduces couplings with the
next neighboring modes with $l\pm2$.  The equations \reff{eq:wIIfora}
therefore comprise two systems of infinite number of equations, the
one with $l=\mathrm{even}$ and the one with $l=\mathrm{odd}$, unless
the background is LRS. When on the other hand the background is LRS,
each equation \reff{eq:wIIfora} for a given $l$ becomes closed itself,
due to the vanishing of the inhomogeneous term $I$.

When the background is LRS, we can find future ($t\goes\infty$)
asymptotic solutions:
\begin{prop}
  On the LRS Bianchi II vacuum solution with $p_1=p_2=2/3$ and
  $p_3=-1/3$, the scalar field equation \reff{eq:wIIfora} for a
  generic mode has the following fundamental solutions as
  $t\goes\infty$:
\begin{equation}
  \label{eq:a_lasym}
  y_l^\pm(t)= 
  t^{-\frac23}e^{\pm i\mu T_{\mathrm{KG}}(t)}(1+o(1)),
\end{equation}
where
\begin{equation}
  T_{\mathrm{KG}}(t)\equiv \frac{9}{16\b}t^{4/3}
      +\frac{\b\mass^2}{\mu^2}t^{2/3}
      +\paren{\frac{3\b}{4}-\frac{8\b^3\mass^4}{27\mu^4}}\log t .
\end{equation}
\end{prop}
The symbol $o(1)$ stands for a function such that
$\lim_{t\goes\infty}o(1)=0$.

\proofmark This result is a generalization of Theorem 3.4,
Ref.\cite{Ta03}, with finite mass $\mass$. As emphasized there, it is
an appropriate choice of new time variable that is essential to obtain
an asymptotic solution. In the present case an appropriate choice
$T(t)$ is given by
\begin{equation}
  \frac{dT}{dt}= 
  \frac{3}{4\b}t^{1/3}
  +\frac{2\b\mass^2}{3\mu^2}t^{-1/3}
  +\paren{\frac{3\b}{4}-\frac{8\b^3\mass^4}{27\mu^4}}\rcp t.
\end{equation}
Following the procedure shown in the reference, one obtains the
asymptotic solution \reff{eq:a_lasym}. \endofproofmark

It is worth noticing that the asymptotic solution only depends on the
fiber index $m$ and the other index $l$ does not affect them. See the
final section for more discussion.

%%%%%%%%%%%%%%%%%%%%
\section{Vector harmonics}
\label{sec:vh}

Let us discuss how we can construct the vector harmonics.

Fortunately, this is on one hand a trivial issue as the invariant
frame $\{\s I,\c I\}$ is well defined on the compactified manifold $M$
(and on the spacetime manifold $M\times\R$).  This means we can define
the components $T_{I\cdots}{}^{J\cdots}$ of any sort of tensor
$T_{a\cdots}{}^{b\cdots}$ with respect to this invariant frame;
\begin{equation}
  T_{I\cdots}{}^{J\cdots}=
  T_{a\cdots}{}^{b\cdots}\cc Ia\cdots\ss Jb\cdots.
\end{equation}
All we have to do is to expand these components with respect to the
scalar harmonics $\phi_{l,m,n_0}$ and $\phi^{(0)}_{k_1,k_2}$. This
procedure corresponds to using the set of one-forms $
\{\phi_{l}\ss1a,\phi_{l}\ss2a,\phi_{l}\ss3a \}_{l=0}^{\infty}$ as
vector harmonics for given $m\neq0$ and $n_0$. (As for $m=0$, $\phi_l$
should of course be replaced by $\phi^{(0)}_{k_1,k_2}$.)  On the
spacetime manifold, we also need the timelike mode vectors, as well as
the replacement $\phi_{l,m,n_0}\goes\varphi_{l,m,n_0}$. We therefore
can define the following harmonics (basis mode vectors)
\begin{equation}
  \vEEd 0la\equiv \varphi_{l}(dt)_a,\;
  \vEEd 1la\equiv \varphi_{l}\ss1a,\;
  \vEEd 2la\equiv \varphi_{l}\ss2a,\;
  \vEEd 3la\equiv \varphi_{l}\ss3a,
\end{equation}
(or the one with $\varphi^{(0)}_{k_1,k_2}$ instead of $\varphi_{l}$
for $m=0$). We call them the \textit{simple vector harmonics}.

The advantage of considering these harmonics is that they are
apparently complete (since the scalar harmonics used are complete).
However, this choice of harmonics does not reflect very well the group
structure regarding the rotational automorphisms \reff{eq:s*}. A more
natural and convenient choice can be obtained by using the ``spherical
bases''\cite{JanEssay} $\AA_I$ ($I=1\sim 3$) defined in
Eqs.\reff{eq:As}, or their duals such that
$\AA_I{}^a\varrho^J{}_a=\delta_I^J$;
\begin{equation}
  \varrho^1\equiv\rcp{\sqrt2}(\s1-\signm i \s2),\quad
  \varrho^2\equiv\rcp{\sqrt2}(\s1+\signm i \s2),\quad
  \varrho^3\equiv \signm i \s3.
\end{equation}
Using these we can set up new harmonics as
\begin{equation}
  \label{eq:defvEE}
  \vEE0la\equiv \varphi_l(dt)_a, \;
  \vEE1la\equiv \varphi_{l+1}\varrho^1{}_a, \;
  \vEE2la\equiv \varphi_{l-1}\varrho^2{}_a, \;
  \vEE3la\equiv \varphi_l\varrho^3{}_a.
\end{equation}
(When $m=0$ we just use the same harmonics as the simple harmonics.)
Beware that $\varphi_{l\pm1}$ are used to define $\vEE1la$ and
$\vEE2la$. The reason will become clear below. Because of this index
correspondence, we should think that the harmonics $\{\vEE
Ila\}_{I=0}^3$ are defined for $l\geq -1$ (not for $l\geq0$). The
basis $\vEE I{-1}a$ for $l=-1$ is nonzero only for $I=1$, and the
others should simply be regarded as zero. We call these harmonics the
\textit{polarized vector harmonics} or the \textit{standard vector
  harmonics}.

We can confirm that for given $m$, the simple harmonics and the
polarized harmonics span the same space of vector fields. This ensures
the completeness of the polarized vector harmonics.
\begin{theorem}
  For given fiber index $m\in\Z$ (and given auxiliary index $n_0$),
  the linear span of the simple harmonics
\begin{equation}
 \mathrm{Span}(V^{\prime m})\equiv
  \brace{ \sum_{l=0}^\infty\sum_{I=0}^{3} c_{I,l}\vEEd Ila 
   \bigg| c_{I,l}\in \mathbf{C} }, \quad (m\neq 0)
\end{equation}
(the $m=0$ case is defined similarly) and that
of the polarized harmonics
\begin{equation}
 \mathrm{Span}(V^{m})\equiv
  \brace{ \sum_{l=-1}^\infty\sum_{I=0}^{3} c_{I,l}\vEE Ila 
   \bigg| c_{I,l}\in \mathbf{C} }, \quad (m\neq 0)
\end{equation}
(the $m=0$ case is defined similarly) are the same;
\begin{equation}
  \mathrm{Span}(V^{\prime m})=\mathrm{Span}(V^{m}).
\end{equation}
\end{theorem}

\proofmark This is trivial for the $m=0$ case, since in this case the
two kinds of harmonics are the same. So, we can assume $m\neq0$. From
the definition \reff{eq:defvEE}, it is apparent that each basis
one-form $\vE Il$ of the polarized harmonics for given $m$ can be
expressed as a linear combination of the simple harmonics belonging to
the same $m$. Conversely, each basis one-form $\vEd Il$ of the
simple harmonics for given $m$ can be expressed as a linear
combination of the polarized harmonics with the same $m$ as
\begin{equation}
  \begin{split}
    \vEd 0l &= \vE 0l, \\
    \vEd 1l &= \rcp{\sqrt2}(\vE 1{l-1}+\vE 2{l+1}), \\
    \vEd 2l &= \frac{\signm i}{\sqrt2}(\vE 1{l-1}-\vE 2{l+1}), \\
    \vEd 3l &= \vE 3l.
  \end{split}
\end{equation}
Therefore the two sets are related by a regular linear transformation,
which proves the identity of the two spans.\endofproofmark

The significance of the polarized harmonics is that for given $l$ (and
as usual, given $m$ and $n_0$), the space spanned by them
\begin{equation}
  \label{eq:SpanV_l}
 \mathrm{Span}(V_l)\equiv \brace{ \sum_{I=0}^{3} c_I\vEE Ila 
   \bigg| c_I\in
\mathbf{C} } 
\end{equation}
is invariant under the operation of $\am^2$. To show this, let us
start with observing the commutation relations
\begin{equation}
  [\AA_1,\AA_2]=\AA_3,\quad 
  [\AA_1,\AA_3]=0,\quad
  [\AA_2,\AA_3]=0,
\end{equation}
from which one can immediately have
\begin{equation}
  \Lie{\AA_I}\AA_J=[\AA_I,\AA_J]=\epsilon_{IJ3}\AA_3,
\end{equation}
where $\epsilon_{IJK}$ is the unit skew symmetric symbol;
$\epsilon_{123}=+1$, $\epsilon_{IJK}=\epsilon_{[IJK]}$. Then, noting
the duality $\AA_J{}^a\varrho^K{}_a=\delta_J^K$, it is also easy to see
\begin{equation}
  \Lie{\AA_I}\varrho^J = -\delta^J_3\epsilon_{IK3}\varrho^K.
\end{equation}
From this equation and the $\chi$-relations, as well as the identity
\begin{equation}
  \am^2=\Lie{\AA_1}\Lie{\AA_2}+\Lie{\AA_2}\Lie{\AA_1},
\end{equation}
one obtains
\begin{equation}
  \am^2(\varphi_l\varrho^I) = -\lambda^2_l\varphi_l\varrho^I
  -2\sqrt{|\mu|}\delta^I_3(\varphi_{l+1}\varrho^1+l\varphi_{l-1}\varrho^2).
\end{equation}
This equation happens to be valid when $I=0$, as well.  Converting to
our bases \reff{eq:defvEE}, we obtain the following statement.
\begin{lemma}
  Let $\vE Il$ ($I=0\sim 3$) be the basis mode one-forms defined in
  Eqs.\reff{eq:defvEE}, and let $\am^2$ be the second order Lie
  derivative operator defined in Eq.\reff{eq:defam^2}. Then, it holds
  that
  \begin{equation}
    \label{eq:am^2vE}
    \begin{split}
    \am^2 \vE0l &= -\lambda^2_{l} \vE0l \\
    \am^2 \vE1l &= -\lambda^2_{l+1} \vE1l \\
    \am^2 \vE2l &= -\lambda^2_{l-1} \vE2l \\
    \am^2 \vE3l &= -\lambda^2_l \vE3l
    -2\sqrt{|\mu|}(\vE1l+l\vE2l),
    \end{split}
  \end{equation}
  where $-\lambda^2_l$ is the eigenvalue of $\am^2$ with respect to
  the mode function $\varphi_l$, defined in
  Eq.\reff{eq:deflambda^2_l}.
\end{lemma}
Note that the right hand sides of Eqs.\reff{eq:am^2vE} are linear
combinations of $\{\vE Il\}_{I=0}^3$ belonging to given $l$. This
proves the invariance we claimed:
\begin{theorem}
  \label{th:invVl}
  The linear span $\mathrm{Span}(V_l)$, defined in
  Eq.\reff{eq:SpanV_l}, of the harmonics $\{\vE Il\}_{I=0}^3$ is
  invariant under the operation of $\am^2$:
\begin{equation}
  \am^2\mathrm{Span}(V_l)\subset \mathrm{Span}(V_l).
\end{equation}
\end{theorem}

Thanks to this property, if the background spacetime is LRS and the
field equation is invariant under the LRS-action induced by $s_\theta$
as well as the usual group action, the ODEs reduced from the field
equation become all independent from the others. In other words,
``each $l$'' decouples from the others in case of LRS. (This property
does not occur for the simple harmonics, since the linear span of them
for a given $l$ is \textit{not} $\am^2$-invariant.)

Before ending this section let us mention the approach taken in
\cite{Ta03}, Sec.3.4, which was based on the analogy of the spherically
symmetric case \cite{RW} or the Bianchi III hyperbolically symmetric
case \cite{TMY}. Let us denote the harmonics used in the reference
with two dashes like $\vEEdd Ila$. $\vEdd 0l$ and $\vEdd 3l$ are
defined in the same way (up to constant multiplication factor) as the
polarized ones;
\begin{equation}
  \label{eq:vEdddef03}
  \vEdd 0l=\vE 0l,\quad \vEdd 3l=\signm\mu \vE 3l.
\end{equation}
$\vEdd 0l$ is called the \textit{time-like basis one-form}, while
$\vEdd 3l$ is called the \textit{fiber basis one-form}. Let us then
consider the ``plane field'' spanned by $\c1$ and $\c2$, which is
horizontal to the base manifold. This horizontal plane field is
invariant under the natural actions of the fibers generated by $\c3$.
We define the ``area two-form'' $\varepsilon$ of this field by $
\varepsilon_{ab}=2\ss1{[a}\ss2{b]}$ (although the plane field is not
integrable). We define the \textit{horizontal gradient (HG) basis
one-form} (corresponding to the ``even'' one-form \cite{RW,TMY}) by
taking gradient of the scalar harmonics and subtracting the fiber
part;
\begin{equation}
  \label{eq:modevS}
  \begin{split}
    \vEEdd 1la &= \del_a\varphi_l-\vEEdd 3la \\
    &= (\c1\varphi_l)\ss1a+(\c2\varphi_l)\ss2a \\
    &= \sqrt{\frac{\abs{\mu}}2}\paren{-
    \paren{\varphi_{l+1}-l\varphi_{l-1}}\ss1a
    +\signm i \paren{\varphi_{l+1}+l\varphi_{l-1}}\ss2a }.
  \end{split}
\end{equation}
And, we define the \textit{dual horizontal gradient (DHG) basis
  one-form} (corresponding to the ``odd'' one-form) by taking the
  dual gradient associated with $\varepsilon$;
\begin{equation}
  \label{eq:modevV}
\begin{split}
  \vEEdd 2la &= i\varepsilon_a{}^b\del_b\varphi_l \\
  &= i((\c2\varphi_l)\ss1a-(\c1\varphi_l)\ss2a) \\
  &= \sqrt{\frac{\abs{\mu}}2}\paren{-\signm
    \paren{\varphi_{l+1}+l\varphi_{l-1}}\ss1a
    + i \paren{\varphi_{l+1}-l\varphi_{l-1}}\ss2a }.
\end{split}
\end{equation}
To raise an index for $\varepsilon_{ab}$ we use the (inverse of the)
standard Bianchi II metric
$h^{(0)ab}=\cc1a\cc1b+\cc2a\cc2b+\cc3a\cc3b$. It is clear that the
harmonics $\harV''_l\equiv\{\vEEdd Ila\}_{I=0}^3$ are equivalent
(i.e., their span is the same) to the polarized harmonics
$\harV_l\equiv\{\vEE Ila\}_{I=0}^3$ \textit{for each} $l$, since they
are merely related to each other by a regular linear transformation,
as seen from the relations
\begin{equation}
  \label{eq:vEdd12}
  \vEdd 1l= -\sqrt{|\mu|}(\vE 1l-l\vE 2l),\quad
  \vEdd 2l= -\signm \sqrt{|\mu|}(\vE 1l+l\vE 2l),
\end{equation}
as well as \reff{eq:vEdddef03}. We call these harmonics the
\textit{mixed vector harmonics}. This choice is particularly
convenient for Maxwell's equation, since the $U(1)$-gauge
transformation, a shifting of the vector potential by a gradient of
scalar, does not affect the component for the DHG basis one-form. (As
a result, this component itself becomes gauge-invariant. See the next
section.)

%%%%%%%%%%%%%%%%%%%%%%%%%
\section{Application to Maxwell's equation}
\label{sec:max}

As an application let us consider the source-free Maxwell equation
$\nabla^aF_{ab}=0$. Since the electromagnetic field $F_{ab}$ is given
by (twice) the exterior derivative of the vector potential $A_a$;
$F_{ab}=\del_a A_b-\del_b A_a$, this equation can be dealt with by
vector harmonics.

Let us consider the irreducible component belonging to given $m\neq0$
and $n_0$. For this component we can expand the vector potential as
follows:
\begin{equation}
  A_a=\sum_{l=-1}^{\infty}\sum_{I=0}^{3} 
  \gamma_I^{(l)}(t)\vEEdd Ila,
\end{equation}
where $\vEEdd Ila$ are the mixed vector harmonics.
The four kinds of functions of time $\gamma_I^{(l)}(t)$ ($I=0\sim 3$)
serve as the field variables.

The quantities we are interested in are the $U(1)$-gauge invariant
variables, which can be easily found by inspecting components of the
field strength $F_{ab}=\del_aA_b-\del_bA_a$. We obtain the following
four independent $U(1)$-invariant variables:
\begin{equation}
  Q_1^{(l)}=\gamma_1^{(l)}-\gamma_3^{(l)},\;
  Q_2^{(l)}=\gamma_2^{(l)},\;
  P_1^{(l)}=\dot\gamma_1^{(l)}-\gamma_0^{(l)},\;
  P_2^{(l)}=\dot\gamma_2^{(l)}.
\end{equation}
Although function $P_3\equiv \dot\gamma_3-\gamma_0$ is also invariant,
it is found that it can be (consistently) solved with the others, due
to the constraint part of Maxwell's equation
$0=(\del_t)^a\nabla^bF_{ab}$, which can be written, using the
invariant operator, as
\begin{equation}
  0= N\inv (q_1\inv\c1\dot A_1+q_2\inv\c2\dot A_2+q_3\inv\c3\dot A_3)-
  \Lap_q A_0,
\end{equation}
where $A_0\equiv N\inv(\del_t)^aA_a$, and $A_I\equiv \cc Ia A_a$
($I=1\sim 3$). Laplacian $\Lap_q$ is given in Eq.\reff{eq:L_q}. 
The evolution equations $0=\cc
Ia\nabla^bF_{ab}$ ($I=1\sim3$) can similarly be written 
\begin{equation}
  \begin{split}
  0 &= -N\inv\dot F_{0I}
  +q_1\inv\c1 F_{1I}+q_2\inv\c2 F_{2I}+q_3\inv\c3 F_{3I}
  + N\inv q_I\inv\dot q_I F_{0I} \\
  & \quad
  -(2N)\inv (q_1\inv\dot q_1+q_2\inv\dot q_2+q_3\inv\dot q_3)F_{0I}
  +(q_1q_2)\inv q_3\delta^3_I F_{12}, \\
  & \quad \text{(no sum for repeated indices)}
  \end{split}
\end{equation}
where $F_{0I}\equiv N\inv(\del_t)^a\cc IbF_{ab}$ and $F_{IJ}\equiv \cc
Ia\cc Jb F_{ab}$. After a rather lengthy computation they become the
following:
\begin{equation}
  \label{eq:modeEMeqns}
  \begin{split}
    \dot Q_1^{(l)} &= P_1^{(l)} +
    \frac{(q_1\inv+q_2\inv)q_3}{2|\mu|} ((2l+1)P_1^{(l)} +\signm P_2^{(l)})
    +I_{Q_1}, \\
    \dot Q_2^{(l)} &= P_2^{(l)}, \\
    \dot P_1^{(l)} &= 
    \paren{\frac{\dot N}{N}-\rcp2\frac{\dot q_3}{q_3}}P_1^{(l)}
    -\mu^2\frac{N^2}{q_3}Q_1^{(l)} +I_{P_1}, \\
    \dot P_2^{(l)} &= \paren{\frac{\dot N}{N}
      -\rcp2\frac{\dot q_3}{q_3}}P_2^{(l)}
    -\mu^2\frac{N^2}{q_3}Q_2^{(l)} \\
    & \quad -\frac{\mu}{2} N^2(q_1\inv+q_2\inv)(Q_1^{(l)}
    +\signm (2l+1)Q_2^{(l)}) +I_{P_2},
  \end{split}
\end{equation}
where the inhomogeneous terms are:
\begin{equation}
  \begin{split}
    I_{Q_1} & \equiv -\frac{(q_1\inv-q_2\inv)q_3}{2|\mu|}\bigg\{
    (l+2)(l+1)(P_1^{(l+2)}-\signm P_2^{(l+2)}) \\
    & \hspace{9.7em} +  (P_1^{(l-2)}+\signm P_2^{(l-2)}) \bigg\},
    \\
    I_{P_1} &\equiv \frac{N^2}{4}(q_1\inv-q_2\inv)|\mu|
    \bigg\{ (l+2)\paren{ Q_1^{(l+2)}+\signm (2l+5) Q_2^{(l+2)} } \\
    & \hspace{9.7em} 
    - l\inv \paren{ Q_1^{(l-2)}+\signm (2l-3) Q_2^{(l-2)} } \bigg\}
    \\
    & \quad - \rcp4 \paren{ \frac{\dot q_1}{q_1}-\frac{\dot q_2}{q_2} }
    \bigg\{ (l+2)\paren{P_1^{(l+2)}-\signm P_2^{(l+2)}} \\
    & \hspace{9.7em} 
    + l\inv \paren{P_1^{(l-2)}+\signm P_2^{(l-2)}} \bigg\}, \\
    I_{P_2} &\equiv \frac{N^2}{4}(q_1\inv-q_2\inv)|\mu|
    \bigg\{ (l+2)\paren{ \signm Q_1^{(l+2)}+ (2l+5) Q_2^{(l+2)} } \\
    & \hspace{9.7em} 
    + l\inv \paren{ \signm Q_1^{(l-2)}+ (2l-3) Q_2^{(l-2)} } \bigg\}
    \\
    & \quad - \rcp4 \paren{ \frac{\dot q_1}{q_1}-\frac{\dot q_2}{q_2} }
    \bigg\{ (l+2)\paren{\signm P_1^{(l+2)}- P_2^{(l+2)}} \\
    & \hspace{9.7em} 
    - l\inv \paren{\signm P_1^{(l-2)}+ P_2^{(l-2)}} \bigg\}.
  \end{split}
\end{equation}

Again, we can see the same qualitative features in the Klein-Gordon
equations; we have two systems of infinite number of equations, the
one with $l=\mathrm{even}$ and the one with $l=\mathrm{odd}$, unless
the background is LRS. When the background is LRS, the couplings
between mode $l$ and the next neighbors $l\pm2$ are cut off due to the
vanishing of the inhomogeneous terms $I_{Q_1}$, $I_{P_1}$ and
$I_{P_2}$ and this makes each system of four first-order equations
\reff{eq:modeEMeqns} for given $l$ closed itself. This is of course a
result of the invariance described in Theorem \ref{th:invVl}.

%%%%%%%%%%%%%%%%%%%%%%
\section{Conclusion}
\label{sec:conc}

There are three main results about the basic properties for the
nilgeometric harmonics obtained in this paper. They are \textbf{(i)}
the irreducible decomposition of the regular representation (Theorem
\ref{th:1}), \textbf{(ii)} the explicit form of the mode functions,
and \textbf{(iii)} the differential representation formula, the
$\chi$-relations (see Eqs.\reff{eq:relc123} and
\reff{eq:relcphi0123}). The decomposition (i) represents the
completeness of our harmonics. As for the point (ii), remember that we
have two kinds of formula, the one for the canonical manifold and the
one for the spacetime manifold. The former is given in
Eqs.\reff{eq:philm2} and \reff{eq:phik1k2}, while the latter is
obtained by the transformations explicitly given in
\S.\ref{sec:gensptm} and \S.\ref{sec:U1-sym}. Remember also that the
$\chi$-relations are the most important for the purpose of separation
of variables. We also have generalized the (scalar) harmonics to
vector harmonics, and demonstrated separation of variables for a
scalar equation (the Klein--Gordon equation) and a vector equation
(the Maxwell equation).

As we have seen, when the fiber index $m$ is nonzero the ODEs reduced
from a field equation, e.g., the ones from the KG equation, become
systems of \textit{infinite number of} simultaneous equations. In this
sense, infinite number of different modes are coupled to each
other. This is a result of the fact that the corresponding irreducible
representation is infinite dimensional. When the background is LRS
however, the couplings between the modes are cut off and as a result,
each single reduced KG equation becomes closed itself. Although the
Maxwell equations give rise to much more complicated reduced equations
because of the multiple components of the field variable, it has the
same feature that the couplings between modes disappear when the
background is LRS. It is also apparent that the linear perturbation
equations will have the same feature if we choose the tensor harmonics
so as to possess the invariance under $\am^2$ like the vector
harmonics do (cf.  Theorem \ref{th:invVl}).

An interesting fact is that as shown in \S.\ref{sec:KG}, the future
asymptotic solution of the LRS KG wave equation depends only on the
fiber index $m$ and does not depend on the spin index $l$. This fact
seems to indicate a clue for analyzing the generic non-LRS cases,
since it suggests that the couplings between the modes asymptotically
disappear even when the background is non-LRS, at least if the
background is close enough to the LRS one. See \cite{Ta04b} for the
same (``fiber term dominated'') behavior of other models (Bianchi VIII
and III). Detailed studies of the non-LRS cases, as well as linear
perturbations of the nilgeometric model, will be reported elsewhere,
on the basis of this work.

\section*{Acknowledgment}

The author wishes to thank Vincent Moncrief for helpful discussions,
especially at an early stage of this work.  He is also grateful to the
organizers, including Akio Hosoya and Tatsuhiko Koike, and the
participants of the \textit{5th Workshop on Singularity, Spacetime and
  Related Physics} held at Keio University, Yokohama, December 19--21
2003, for the kind invitation and many beneficial comments.


\begin{thebibliography}{AA}
\bibitem{note1} The ``left'' regular representation can also be
  defined similarly. We take the ``right'' one for the reason
  explained below.
\bibitem{note3} While $\Isom\N$
  is four-dimensional, the extra isometries are not necessary for us
  to represent $\mathbf{a}_i$, resulting in $\mathbf{a}_i\in\G{II}$
  here.  Remark however that to obtain this representation the fourth
  dimensional isometries play an essential role, especially to delete
  the first component of $\mathbf{a}_2$.
\bibitem{note5} We could impose a unitary condition on
  $\AA_i$ $(i=1,2)$ to make the choice of $\alpha_l$ and $\beta_l$
  unique, although there is no reason to do so for our purpose. Our
  present choice \reff{eq:atmp} makes the final formula (see
  Eqs.\reff{eq:relc123}) to some extent simpler, with less `square
  roots' appearing in the formula than in the one obtained requiring
  the unitary condition.  It is however easy to convert the resulting
  mode functions into those for unitary $\AA_i$. See the last
  paragraph of \S.\ref{sec:mf}.
\bibitem{note6}  We are assuming that the
  action of $A$ on the spacetime manifold $\tilde M\times\R$ is
  defined by
\begin{equation}
  A\circ (\x,t)=(A\circ\x,t),\quad (\x,t)\in\tilde M\times\R.
\end{equation}
(Apparently, the same equality for $\Gamma$ defines its action on
$\tilde M$.)
\bibitem{note7} The parameterization among the free
  parameters has been rearranged from \cite{TKH1} so that it is
  convenient for our purpose.
\end{thebibliography}
\end{document}